\pdfoutput=1
\documentclass[twocolumn]{aastex6}

\usepackage{times}
\usepackage{graphicx}
\usepackage{amssymb}
\usepackage{amsmath}
\usepackage{pifont}
\usepackage{graphics}
\usepackage{xcolor}
\usepackage{epsfig}
\usepackage{verbatim}
\usepackage{booktabs}

\newcommand{\bjdtdb}{BJD$_{\rm TDB}\ $}      
\newcommand{\feh}{\ensuremath{\left[{\rm Fe}/{\rm H}\right]}}  
\newcommand{\meh}{\ensuremath{\left[{\rm m}/{\rm H}\right]}} 
\newcommand{\ecosw}{\ensuremath{e\cos{\omega_\star}}} 
\newcommand{\esinw}{\ensuremath{e\sin{\omega_\star}}}
\newcommand{\teff}{\ensuremath{T_{\rm eff}}}
\newcommand{\loggstar}{\ensuremath{\log{g_*}}}
\newcommand{\vsinistar}{\ensuremath{v\sin{i}}}
\newcommand{\msun}{\ensuremath{\,M_\Sun}}
\newcommand{\rsun}{\ensuremath{\,R_\Sun}}
\newcommand{\lsun}{\ensuremath{\,L_\Sun}}
\newcommand{\mj}{\ensuremath{\,M_{\rm J}}}
\newcommand{\rj}{\ensuremath{\,R_{\rm J}}}
\newcommand{\fave}{\langle F \rangle}
\newcommand{\fluxcgs}{\ensuremath{\rm 10^9 erg~s^{-1} cm^{-2}}}
\newcommand{\densitycgs}{\ensuremath{\rm g~cm^{-3}}}
\newcommand{\gravitycgs}{\ensuremath{\rm cm~s^{-2}}}

\newcommand{\kms}{\ensuremath{\rm km\ s^{-1}}}

\newcommand{\mstarvalone}{\ensuremath{1.550_{-0.081}^{+0.083}}}
\newcommand{\mstarvaltwo}{\ensuremath{1.524_{-0.068}^{+0.069}}}
\newcommand{\mstarvalthree}{\ensuremath{1.513_{-0.085}^{+0.090}}}
\newcommand{\mstarvalfour}{\ensuremath{1.490_{-0.080}^{+0.083}}}
\newcommand{\rstarvalone}{\ensuremath{1.98_{-0.11}^{+0.19}}}
\newcommand{\rstarvaltwo}{\ensuremath{1.908_{-0.035}^{+0.042}}}
\newcommand{\rstarvalthree}{\ensuremath{1.98_{-0.11}^{+0.19}}}
\newcommand{\rstarvalfour}{\ensuremath{1.895_{-0.040}^{+0.046}}}
\newcommand{\lstarvalone}{\ensuremath{7.02_{-0.99}^{+1.4}}}
\newcommand{\lstarvaltwo}{\ensuremath{6.50_{-0.58}^{+0.65}}}
\newcommand{\lstarvalthree}{\ensuremath{7.02_{-0.99}^{+1.4}}}
\newcommand{\lstarvalfour}{\ensuremath{6.42_{-0.57}^{+0.64}}}
\newcommand{\rhovalone}{\ensuremath{0.282_{-0.060}^{+0.041}}}
\newcommand{\rhovaltwo}{\ensuremath{0.3111_{-0.014}^{+0.0070}}}
\newcommand{\rhovalthree}{\ensuremath{0.277_{-0.060}^{+0.042}}}
\newcommand{\rhovalfour}{\ensuremath{0.3111_{-0.014}^{+0.0070}}}
\newcommand{\loggstarvalone}{\ensuremath{4.034_{-0.064}^{+0.038}}}
\newcommand{\loggstarvaltwo}{\ensuremath{4.0599_{-0.014}^{+0.0096}}}
\newcommand{\loggstarvalthree}{\ensuremath{4.026_{-0.067}^{+0.041}}}
\newcommand{\loggstarvalfour}{\ensuremath{4.057_{-0.014}^{+0.011}}}
\newcommand{\teffvalone}{\ensuremath{6670\pm120}}
\newcommand{\teffvaltwo}{\ensuremath{6670\pm120}}
\newcommand{\teffvalthree}{\ensuremath{6670\pm120}}
\newcommand{\teffvalfour}{\ensuremath{6680\pm110}}
\newcommand{\fehvalone}{\ensuremath{0.08\pm0.13}}
\newcommand{\fehvaltwo}{\ensuremath{0.09\pm0.13}}
\newcommand{\fehvalthree}{\ensuremath{0.08\pm0.13}}
\newcommand{\fehvalfour}{\ensuremath{0.08\pm0.13}}
\newcommand{\evalone}{\ensuremath{0.058_{-0.042}^{+0.071}}}
\newcommand{\evaltwo}{\ensuremath{--}}
\newcommand{\evalthree}{\ensuremath{0.061_{-0.044}^{+0.074}}}
\newcommand{\evalfour}{\ensuremath{--}}
\newcommand{\omegavalone}{\ensuremath{106_{-83}^{+73}}}
\newcommand{\omegavaltwo}{\ensuremath{--}}
\newcommand{\omegavalthree}{\ensuremath{105_{-77}^{+63}}}
\newcommand{\omegavalfour}{\ensuremath{--}}
\newcommand{\pvalone}{\ensuremath{2.8717518\pm0.0000029}}
\newcommand{\pvaltwo}{\ensuremath{2.8717518\pm0.0000028}}
\newcommand{\pvalthree}{\ensuremath{2.8717518\pm0.0000029}}
\newcommand{\pvalfour}{\ensuremath{2.8717518\pm0.0000029}}
\newcommand{\avalone}{\ensuremath{0.04576_{-0.00081}^{+0.00080}}}
\newcommand{\avaltwo}{\ensuremath{0.04550_{-0.00069}^{+0.00067}}}
\newcommand{\avalthree}{\ensuremath{0.04539_{-0.00087}^{+0.00088}}}
\newcommand{\avalfour}{\ensuremath{0.04517\pm0.00082}}
\newcommand{\mpvalone}{\ensuremath{1.18\pm0.13}}
\newcommand{\mpvaltwo}{\ensuremath{1.18\pm0.11}}
\newcommand{\mpvalthree}{\ensuremath{1.17\pm0.13}}
\newcommand{\mpvalfour}{\ensuremath{1.16\pm0.11}}
\newcommand{\rpvalone}{\ensuremath{1.628_{-0.093}^{+0.15}}}
\newcommand{\rpvaltwo}{\ensuremath{1.570_{-0.036}^{+0.042}}}
\newcommand{\rpvalthree}{\ensuremath{1.627_{-0.096}^{+0.15}}}
\newcommand{\rpvalfour}{\ensuremath{1.561_{-0.039}^{+0.045}}}
\newcommand{\rhopvalone}{\ensuremath{0.335_{-0.076}^{+0.074}}}
\newcommand{\rhopvaltwo}{\ensuremath{0.377\pm0.040}}
\newcommand{\rhopvalthree}{\ensuremath{0.331_{-0.076}^{+0.074}}}
\newcommand{\rhopvalfour}{\ensuremath{0.377_{-0.041}^{+0.042}}}
\newcommand{\loggpvalone}{\ensuremath{3.037_{-0.077}^{+0.066}}}
\newcommand{\loggpvaltwo}{\ensuremath{3.073_{-0.044}^{+0.040}}}
\newcommand{\loggpvalthree}{\ensuremath{3.032_{-0.078}^{+0.067}}}
\newcommand{\loggpvalfour}{\ensuremath{3.070_{-0.045}^{+0.041}}}
\newcommand{\teqvalone}{\ensuremath{2120_{-66}^{+87}}}
\newcommand{\teqvaltwo}{\ensuremath{2085_{-38}^{+39}}}
\newcommand{\teqvalthree}{\ensuremath{2127_{-67}^{+89}}}
\newcommand{\teqvalfour}{\ensuremath{2085_{-37}^{+39}}}
\newcommand{\thetavalone}{\ensuremath{0.0425_{-0.0053}^{+0.0055}}}
\newcommand{\thetavaltwo}{\ensuremath{0.0448\pm0.0042}}
\newcommand{\thetavalthree}{\ensuremath{0.0426_{-0.0053}^{+0.0055}}}
\newcommand{\thetavalfour}{\ensuremath{0.0449_{-0.0042}^{+0.0043}}}
\newcommand{\favevalone}{\ensuremath{4.57_{-0.53}^{+0.73}}}
\newcommand{\favevaltwo}{\ensuremath{4.29_{-0.30}^{+0.33}}}
\newcommand{\favevalthree}{\ensuremath{4.63_{-0.55}^{+0.76}}}
\newcommand{\favevalfour}{\ensuremath{4.29_{-0.30}^{+0.33}}}
\newcommand{\tcvalone}{\ensuremath{2457542.52505\pm0.00040}}
\newcommand{\tcvaltwo}{\ensuremath{2457542.52504\pm0.00039}}
\newcommand{\tcvalthree}{\ensuremath{2457542.52504\pm0.00040}}
\newcommand{\tcvalfour}{\ensuremath{2457542.52504\pm0.00040}}
\newcommand{\tpvalone}{\ensuremath{2457542.62_{-0.64}^{+0.53}}}
\newcommand{\tpvaltwo}{\ensuremath{--}}
\newcommand{\tpvalthree}{\ensuremath{2457542.62_{-0.56}^{+0.47}}}
\newcommand{\tpvalfour}{\ensuremath{--}}
\newcommand{\kvalone}{\ensuremath{127\pm13}}
\newcommand{\kvaltwo}{\ensuremath{127\pm11}}
\newcommand{\kvalthree}{\ensuremath{127\pm13}}
\newcommand{\kvalfour}{\ensuremath{127_{-11}^{+12}}}
\newcommand{\mpsinivalone}{\ensuremath{1.18\pm0.13}}
\newcommand{\mpsinivaltwo}{\ensuremath{1.18\pm0.11}}
\newcommand{\mpsinivalthree}{\ensuremath{1.17\pm0.13}}
\newcommand{\mpsinivalfour}{\ensuremath{1.16\pm0.11}}
\newcommand{\mpmstarvalone}{\ensuremath{0.000731_{-0.000076}^{+0.000077}}}
\newcommand{\mpmstarvaltwo}{\ensuremath{0.000740_{-0.000067}^{+0.000068}}}
\newcommand{\mpmstarvalthree}{\ensuremath{0.000738\pm0.000077}}
\newcommand{\mpmstarvalfour}{\ensuremath{0.000744\pm0.000069}}
\newcommand{\uvalone}{\ensuremath{0.5778_{-0.0074}^{+0.0078}}}
\newcommand{\uvaltwo}{\ensuremath{0.5779_{-0.0073}^{+0.0078}}}
\newcommand{\uvalthree}{\ensuremath{0.5777_{-0.0073}^{+0.0078}}}
\newcommand{\uvalfour}{\ensuremath{0.5776_{-0.0072}^{+0.0077}}}
\newcommand{\gammaapfvalone}{\ensuremath{-27\pm30}}
\newcommand{\gammaapfvaltwo}{\ensuremath{-26\pm24}}
\newcommand{\gammaapfvalthree}{\ensuremath{-28\pm30}}
\newcommand{\gammaapfvalfour}{\ensuremath{-27\pm24}}
\newcommand{\gammatresvalone}{\ensuremath{-57.1\pm9.7}}
\newcommand{\gammatresvaltwo}{\ensuremath{-57.4\pm8.7}}
\newcommand{\gammatresvalthree}{\ensuremath{-56.9\pm9.6}}
\newcommand{\gammatresvalfour}{\ensuremath{-57.3\pm8.7}}
\newcommand{\dotgammavalone}{\ensuremath{-0.39_{-0.68}^{+0.67}}}
\newcommand{\dotgammavaltwo}{\ensuremath{-0.39\pm0.58}}
\newcommand{\dotgammavalthree}{\ensuremath{-0.40\pm0.68}}
\newcommand{\dotgammavalfour}{\ensuremath{-0.40\pm0.58}}
\newcommand{\ecoswvalone}{\ensuremath{-0.006_{-0.051}^{+0.032}}}
\newcommand{\ecoswvaltwo}{\ensuremath{--}}
\newcommand{\ecoswvalthree}{\ensuremath{-0.007_{-0.052}^{+0.032}}}
\newcommand{\ecoswvalfour}{\ensuremath{--}}
\newcommand{\esinwvalone}{\ensuremath{0.027_{-0.041}^{+0.082}}}
\newcommand{\esinwvaltwo}{\ensuremath{--}}
\newcommand{\esinwvalthree}{\ensuremath{0.033_{-0.043}^{+0.083}}}
\newcommand{\esinwvalfour}{\ensuremath{--}}
\newcommand{\fmonemtwovalone}{\ensuremath{0.00000063_{-0.00000018}^{+0.00000022}}}
\newcommand{\fmonemtwovaltwo}{\ensuremath{0.00000064_{-0.00000016}^{+0.00000019}}}
\newcommand{\fmonemtwovalthree}{\ensuremath{0.00000064_{-0.00000018}^{+0.00000022}}}
\newcommand{\fmonemtwovalfour}{\ensuremath{0.00000064_{-0.00000016}^{+0.00000019}}}
\newcommand{\rprstarvalone}{\ensuremath{0.08462\pm0.00091}}
\newcommand{\rprstarvaltwo}{\ensuremath{0.08462\pm0.00091}}
\newcommand{\rprstarvalthree}{\ensuremath{0.08462\pm0.00091}}
\newcommand{\rprstarvalfour}{\ensuremath{0.08471_{-0.00090}^{+0.00091}}}
\newcommand{\arvalone}{\ensuremath{4.97_{-0.38}^{+0.23}}}
\newcommand{\arvaltwo}{\ensuremath{5.138_{-0.078}^{+0.038}}}
\newcommand{\arvalthree}{\ensuremath{4.94_{-0.39}^{+0.24}}}
\newcommand{\arvalfour}{\ensuremath{5.138_{-0.079}^{+0.038}}}
\newcommand{\ivalone}{\ensuremath{88.80_{-1.3}^{+0.84}}}
\newcommand{\ivaltwo}{\ensuremath{88.86_{-1.2}^{+0.79}}}
\newcommand{\ivalthree}{\ensuremath{88.76_{-1.3}^{+0.87}}}
\newcommand{\ivalfour}{\ensuremath{88.85_{-1.2}^{+0.80}}}
\newcommand{\bvalone}{\ensuremath{0.099_{-0.069}^{+0.10}}}
\newcommand{\bvaltwo}{\ensuremath{0.102_{-0.071}^{+0.10}}}
\newcommand{\bvalthree}{\ensuremath{0.101_{-0.071}^{+0.10}}}
\newcommand{\bvalfour}{\ensuremath{0.103_{-0.072}^{+0.10}}}
\newcommand{\deltavalone}{\ensuremath{0.00716_{-0.00015}^{+0.00016}}}
\newcommand{\deltavaltwo}{\ensuremath{0.00716\pm0.00015}}
\newcommand{\deltavalthree}{\ensuremath{0.00716\pm0.00015}}
\newcommand{\deltavalfour}{\ensuremath{0.00718\pm0.00015}}
\newcommand{\tfwhmvalone}{\ensuremath{0.17783_{-0.00086}^{+0.00087}}}
\newcommand{\tfwhmvaltwo}{\ensuremath{0.17792_{-0.00085}^{+0.00086}}}
\newcommand{\tfwhmvalthree}{\ensuremath{0.17783\pm0.00087}}
\newcommand{\tfwhmvalfour}{\ensuremath{0.17792_{-0.00085}^{+0.00086}}}
\newcommand{\tauvalone}{\ensuremath{0.01545_{-0.00028}^{+0.00052}}}
\newcommand{\tauvaltwo}{\ensuremath{0.01545_{-0.00028}^{+0.00055}}}
\newcommand{\tauvalthree}{\ensuremath{0.01546_{-0.00028}^{+0.00053}}}
\newcommand{\tauvalfour}{\ensuremath{0.01547_{-0.00028}^{+0.00056}}}
\newcommand{\tonefourvalone}{\ensuremath{0.1934_{-0.0010}^{+0.0011}}}
\newcommand{\tonefourvaltwo}{\ensuremath{0.1935_{-0.0010}^{+0.0011}}}
\newcommand{\tonefourvalthree}{\ensuremath{0.1934_{-0.0010}^{+0.0011}}}
\newcommand{\tonefourvalfour}{\ensuremath{0.1935_{-0.0010}^{+0.0011}}}
\newcommand{\ptvalone}{\ensuremath{0.189_{-0.015}^{+0.035}}}
\newcommand{\ptvaltwo}{\ensuremath{0.1782_{-0.0013}^{+0.0027}}}
\newcommand{\ptvalthree}{\ensuremath{0.192_{-0.016}^{+0.036}}}
\newcommand{\ptvalfour}{\ensuremath{0.1782_{-0.0013}^{+0.0027}}}
\newcommand{\ptgvalone}{\ensuremath{0.224_{-0.018}^{+0.041}}}
\newcommand{\ptgvaltwo}{\ensuremath{0.2111_{-0.0016}^{+0.0033}}}
\newcommand{\ptgvalthree}{\ensuremath{0.227_{-0.019}^{+0.043}}}
\newcommand{\ptgvalfour}{\ensuremath{0.2111_{-0.0016}^{+0.0034}}}

\newcommand{\tczerovalone}{\ensuremath{2457493.70450_{-0.00085}^{+0.00082}}}
\newcommand{\tczerovaltwo}{\ensuremath{2457493.70451_{-0.00084}^{+0.00082}}}
\newcommand{\tczerovalthree}{\ensuremath{2457493.70449_{-0.00085}^{+0.00083}}}
\newcommand{\tczerovalfour}{\ensuremath{2457493.70453_{-0.00084}^{+0.00081}}}
\newcommand{\tconevalone}{\ensuremath{2457493.7064\pm0.0011}}
\newcommand{\tconevaltwo}{\ensuremath{2457493.7064\pm0.0011}}
\newcommand{\tconevalthree}{\ensuremath{2457493.7064\pm0.0011}}
\newcommand{\tconevalfour}{\ensuremath{2457493.7065\pm0.0011}}
\newcommand{\tctwovalone}{\ensuremath{2457493.70459_{-0.00087}^{+0.00086}}}
\newcommand{\tctwovaltwo}{\ensuremath{2457493.70460_{-0.00087}^{+0.00086}}}
\newcommand{\tctwovalthree}{\ensuremath{2457493.70459\pm0.00087}}
\newcommand{\tctwovalfour}{\ensuremath{2457493.70458_{-0.00087}^{+0.00086}}}
\newcommand{\tcthreevalone}{\ensuremath{2457496.5787_{-0.0018}^{+0.0017}}}
\newcommand{\tcthreevaltwo}{\ensuremath{2457496.5787_{-0.0018}^{+0.0017}}}
\newcommand{\tcthreevalthree}{\ensuremath{2457496.5787_{-0.0018}^{+0.0017}}}
\newcommand{\tcthreevalfour}{\ensuremath{2457496.5787_{-0.0018}^{+0.0017}}}
\newcommand{\tcfourvalone}{\ensuremath{2457539.6551\pm0.0017}}
\newcommand{\tcfourvaltwo}{\ensuremath{2457539.6551\pm0.0017}}
\newcommand{\tcfourvalthree}{\ensuremath{2457539.6551\pm0.0017}}
\newcommand{\tcfourvalfour}{\ensuremath{2457539.6551\pm0.0017}}
\newcommand{\tcfivevalone}{\ensuremath{2457545.3962\pm0.0011}}
\newcommand{\tcfivevaltwo}{\ensuremath{2457545.3962\pm0.0011}}
\newcommand{\tcfivevalthree}{\ensuremath{2457545.3962\pm0.0011}}
\newcommand{\tcfivevalfour}{\ensuremath{2457545.3963\pm0.0011}}
\newcommand{\tcsixvalone}{\ensuremath{2457559.7568\pm0.0011}}
\newcommand{\tcsixvaltwo}{\ensuremath{2457559.7568\pm0.0011}}
\newcommand{\tcsixvalthree}{\ensuremath{2457559.7568\pm0.0011}}
\newcommand{\tcsixvalfour}{\ensuremath{2457559.7568\pm0.0011}}
\newcommand{\tcsevenvalone}{\ensuremath{2457559.7572\pm0.0020}}
\newcommand{\tcsevenvaltwo}{\ensuremath{2457559.7572\pm0.0020}}
\newcommand{\tcsevenvalthree}{\ensuremath{2457559.7572_{-0.0021}^{+0.0020}}}
\newcommand{\tcsevenvalfour}{\ensuremath{2457559.7572\pm0.0020}}
\newcommand{\tceightvalone}{\ensuremath{2457559.7536_{-0.0020}^{+0.0019}}}
\newcommand{\tceightvaltwo}{\ensuremath{2457559.7536_{-0.0020}^{+0.0019}}}
\newcommand{\tceightvalthree}{\ensuremath{2457559.7536_{-0.0020}^{+0.0019}}}
\newcommand{\tceightvalfour}{\ensuremath{2457559.7536_{-0.0020}^{+0.0019}}}
\newcommand{\tcninevalone}{\ensuremath{2457588.4708_{-0.0013}^{+0.0014}}}
\newcommand{\tcninevaltwo}{\ensuremath{2457588.4709_{-0.0013}^{+0.0014}}}
\newcommand{\tcninevalthree}{\ensuremath{2457588.4708_{-0.0013}^{+0.0014}}}
\newcommand{\tcninevalfour}{\ensuremath{2457588.4709_{-0.0013}^{+0.0014}}}
\newcommand{\tconezerovalone}{\ensuremath{2457591.3461_{-0.0016}^{+0.0015}}}
\newcommand{\tconezerovaltwo}{\ensuremath{2457591.3461_{-0.0016}^{+0.0015}}}
\newcommand{\tconezerovalthree}{\ensuremath{2457591.3461_{-0.0016}^{+0.0015}}}
\newcommand{\tconezerovalfour}{\ensuremath{2457591.3461_{-0.0016}^{+0.0015}}}
\newcommand{\uoneivalone}{\ensuremath{0.1792_{-0.0098}^{+0.010}}}
\newcommand{\uoneivaltwo}{\ensuremath{0.1797_{-0.0097}^{+0.0100}}}
\newcommand{\uoneivalthree}{\ensuremath{0.1789_{-0.0096}^{+0.010}}}
\newcommand{\uoneivalfour}{\ensuremath{0.1795_{-0.0094}^{+0.0098}}}
\newcommand{\utwoivalone}{\ensuremath{0.3237_{-0.0070}^{+0.0064}}}
\newcommand{\utwoivaltwo}{\ensuremath{0.3236_{-0.0071}^{+0.0062}}}
\newcommand{\utwoivalthree}{\ensuremath{0.3238_{-0.0069}^{+0.0064}}}
\newcommand{\utwoivalfour}{\ensuremath{0.3233_{-0.0069}^{+0.0063}}}
\newcommand{\uonervalone}{\ensuremath{0.250\pm0.010}}
\newcommand{\uonervaltwo}{\ensuremath{0.251\pm0.010}}
\newcommand{\uonervalthree}{\ensuremath{0.2502_{-0.0099}^{+0.010}}}
\newcommand{\uonervalfour}{\ensuremath{0.2504_{-0.0097}^{+0.010}}}
\newcommand{\utworvalone}{\ensuremath{0.3343_{-0.0065}^{+0.0067}}}
\newcommand{\utworvaltwo}{\ensuremath{0.3344\pm0.0066}}
\newcommand{\utworvalthree}{\ensuremath{0.3344_{-0.0065}^{+0.0066}}}
\newcommand{\utworvalfour}{\ensuremath{0.3342_{-0.0064}^{+0.0065}}}
\newcommand{\uonesloangvalone}{\ensuremath{0.422_{-0.014}^{+0.015}}}
\newcommand{\uonesloangvaltwo}{\ensuremath{0.422_{-0.014}^{+0.015}}}
\newcommand{\uonesloangvalthree}{\ensuremath{0.422_{-0.014}^{+0.015}}}
\newcommand{\uonesloangvalfour}{\ensuremath{0.421_{-0.013}^{+0.015}}}
\newcommand{\utwosloangvalone}{\ensuremath{0.3013_{-0.0090}^{+0.0095}}}
\newcommand{\utwosloangvaltwo}{\ensuremath{0.3018_{-0.0090}^{+0.0093}}}
\newcommand{\utwosloangvalthree}{\ensuremath{0.3013_{-0.0090}^{+0.0093}}}
\newcommand{\utwosloangvalfour}{\ensuremath{0.3019_{-0.0088}^{+0.0090}}}
\newcommand{\uonesloanivalone}{\ensuremath{0.1977_{-0.0098}^{+0.0100}}}
\newcommand{\uonesloanivaltwo}{\ensuremath{0.1982_{-0.0097}^{+0.0100}}}
\newcommand{\uonesloanivalthree}{\ensuremath{0.1974_{-0.0096}^{+0.0100}}}
\newcommand{\uonesloanivalfour}{\ensuremath{0.1979_{-0.0094}^{+0.0098}}}
\newcommand{\utwosloanivalone}{\ensuremath{0.3259_{-0.0071}^{+0.0067}}}
\newcommand{\utwosloanivaltwo}{\ensuremath{0.3259_{-0.0072}^{+0.0065}}}
\newcommand{\utwosloanivalthree}{\ensuremath{0.3260_{-0.0070}^{+0.0067}}}
\newcommand{\utwosloanivalfour}{\ensuremath{0.3256_{-0.0070}^{+0.0066}}}
\newcommand{\uonesloanrvalone}{\ensuremath{0.272\pm0.010}}
\newcommand{\uonesloanrvaltwo}{\ensuremath{0.272\pm0.010}}
\newcommand{\uonesloanrvalthree}{\ensuremath{0.2715_{-0.0099}^{+0.010}}}
\newcommand{\uonesloanrvalfour}{\ensuremath{0.2716_{-0.0097}^{+0.010}}}
\newcommand{\utwosloanrvalone}{\ensuremath{0.3350_{-0.0063}^{+0.0066}}}
\newcommand{\utwosloanrvaltwo}{\ensuremath{0.3352_{-0.0063}^{+0.0065}}}
\newcommand{\utwosloanrvalthree}{\ensuremath{0.3351_{-0.0062}^{+0.0065}}}
\newcommand{\utwosloanrvalfour}{\ensuremath{0.3350_{-0.0062}^{+0.0064}}}
\newcommand{\uonevvalone}{\ensuremath{0.337_{-0.010}^{+0.011}}}
\newcommand{\uonevvaltwo}{\ensuremath{0.337_{-0.010}^{+0.011}}}
\newcommand{\uonevvalthree}{\ensuremath{0.337_{-0.010}^{+0.011}}}
\newcommand{\uonevvalfour}{\ensuremath{0.3365_{-0.0099}^{+0.011}}}
\newcommand{\utwovvalone}{\ensuremath{0.3227_{-0.0059}^{+0.0067}}}
\newcommand{\utwovvaltwo}{\ensuremath{0.3229_{-0.0059}^{+0.0066}}}
\newcommand{\utwovvalthree}{\ensuremath{0.3227_{-0.0058}^{+0.0065}}}
\newcommand{\utwovvalfour}{\ensuremath{0.3228_{-0.0057}^{+0.0064}}}
\newcommand{\tsvalone}{\ensuremath{2457543.950_{-0.094}^{+0.058}}}
\newcommand{\tsvaltwo}{\ensuremath{2457541.08916\pm0.00039}}
\newcommand{\tsvalthree}{\ensuremath{2457543.948_{-0.095}^{+0.059}}}
\newcommand{\tsvalfour}{\ensuremath{2457541.08917\pm0.00040}}
\newcommand{\bsvalone}{\ensuremath{0.107_{-0.075}^{+0.11}}}
\newcommand{\bsvaltwo}{\ensuremath{--}}
\newcommand{\bsvalthree}{\ensuremath{0.111_{-0.077}^{+0.11}}}
\newcommand{\bsvalfour}{\ensuremath{--}}
\newcommand{\tsfwhmvalone}{\ensuremath{0.188_{-0.014}^{+0.032}}}
\newcommand{\tsfwhmvaltwo}{\ensuremath{--}}
\newcommand{\tsfwhmvalthree}{\ensuremath{0.190_{-0.015}^{+0.033}}}
\newcommand{\tsfwhmvalfour}{\ensuremath{--}}
\newcommand{\tausvalone}{\ensuremath{0.0165_{-0.0015}^{+0.0030}}}
\newcommand{\tausvaltwo}{\ensuremath{--}}
\newcommand{\tausvalthree}{\ensuremath{0.0167_{-0.0016}^{+0.0031}}}
\newcommand{\tausvalfour}{\ensuremath{--}}
\newcommand{\tsonefourvalone}{\ensuremath{0.204_{-0.016}^{+0.035}}}
\newcommand{\tsonefourvaltwo}{\ensuremath{--}}
\newcommand{\tsonefourvalthree}{\ensuremath{0.207_{-0.017}^{+0.036}}}
\newcommand{\tsonefourvalfour}{\ensuremath{--}}
\newcommand{\psvalone}{\ensuremath{0.1790_{-0.0017}^{+0.0031}}}
\newcommand{\psvaltwo}{\ensuremath{--}}
\newcommand{\psvalthree}{\ensuremath{0.1791_{-0.0017}^{+0.0032}}}
\newcommand{\psvalfour}{\ensuremath{--}}
\newcommand{\psgvalone}{\ensuremath{0.2121_{-0.0020}^{+0.0038}}}
\newcommand{\psgvaltwo}{\ensuremath{--}}
\newcommand{\psgvalthree}{\ensuremath{0.2122_{-0.0021}^{+0.0039}}}
\newcommand{\psgvalfour}{\ensuremath{--}}

\begin{document}
\title{KELT-18\MakeLowercase{b}: Puffy Planet, Hot Host, Probably Perturbed}

\author{
Kim K. McLeod\altaffilmark{1},
Joseph E. Rodriguez\altaffilmark{2},
Ryan J. Oelkers\altaffilmark{3},
Karen A. Collins\altaffilmark{3},
Allyson  Bieryla\altaffilmark{2},
Benjamin J. Fulton\altaffilmark{4,5},
Keivan G. Stassun\altaffilmark{3,6},
B. Scott  Gaudi\altaffilmark{7},
Kaloyan  Penev\altaffilmark{8},
Daniel J. Stevens\altaffilmark{7},
Knicole D. Col\'on\altaffilmark{9,10,44},
Joshua  Pepper\altaffilmark{11},
Norio Narita\altaffilmark{12,13,14},
Ryu Tsuguru\altaffilmark{14,15},
Akihiko Fukui\altaffilmark{16},
Phillip A. Reed\altaffilmark{17},
Bethany Tirrell\altaffilmark{17},
Tiffany Visgaitis\altaffilmark{17},
John F. Kielkopf\altaffilmark{18},
David H. Cohen\altaffilmark{19},
Eric L. N. Jensen\altaffilmark{19},
Joao Gregorio\altaffilmark{20},
\"{O}zg\"{u}r Ba\c{s}t\"{u}rk\altaffilmark{21},
Thomas E. Oberst\altaffilmark{22},
Casey Melton\altaffilmark{1},
Eliza M.-R. Kempton\altaffilmark{23},
Andrew Baldrige\altaffilmark{23},
Y. Sunny Zhao\altaffilmark{23},
Roberto Zambelli\altaffilmark{24},
David W. Latham\altaffilmark{2}, 
Gilbert A. Esquerdo\altaffilmark{2}, 
Perry Berlind\altaffilmark{2}, 
Michael L. Calkins\altaffilmark{2}, 
Andrew W. Howard\altaffilmark{25},  
Howard Isaacson\altaffilmark{26},  
Lauren M. Weiss\altaffilmark{27,28}, 
Paul Benni\altaffilmark{29},
Thomas G. Beatty\altaffilmark{30,31},
Jason D. Eastman\altaffilmark{2},
Matthew T. Penny\altaffilmark{7,32},
Robert J. Siverd\altaffilmark{33},
Michael B. Lund\altaffilmark{3},
Jonathan  Labadie-Bartz\altaffilmark{11},
G. Zhou\altaffilmark{2},   
Ivan A. Curtis\altaffilmark{34},
Michael D. Joner\altaffilmark{35},
Mark  Manner\altaffilmark{36},
Howard Relles\altaffilmark{2}, 
Gaetano Scarpetta\altaffilmark{37,38},  
Denise C. Stephens\altaffilmark{35},
Chris Stockdale\altaffilmark{39},
T.G. Tan\altaffilmark{40},
D. L. DePoy\altaffilmark{41,42},
Jennifer L. Marshall\altaffilmark{41,42},
Richard W. Pogge\altaffilmark{7},
Mark Trueblood\altaffilmark{43},  
\and Patricia  Trueblood\altaffilmark{43}
}

\altaffiltext{1}{Department of Astronomy, Wellesley College, Wellesley, MA 02481, USA; kmcleod@wellesley.edu}
\altaffiltext{2}{Harvard-Smithsonian Center for Astrophysics, Cambridge, MA 02138, USA}
\altaffiltext{3}{Department of Physics and Astronomy, Vanderbilt University, Nashville, TN 37235, USA}
\altaffiltext{4}{Institute for Astronomy, University of Hawaii, Honolulu, HI 96822, USA}
\altaffiltext{5}{NSF Graduate Research Fellow}
\altaffiltext{6}{Department of Physics, Fisk University, Nashville, TN 37208, USA}
\altaffiltext{7}{Department of Astronomy, The Ohio State University, Columbus, OH 43210, USA}
\altaffiltext{8}{Department of Physics, Fisk University, Nashville, TN 37208, USA}
\altaffiltext{9}{NASA Ames Research Center, M/S 244-30, Moffett Field, CA 94035, USA}
\altaffiltext{10}{Bay Area Environmental Research Institute, Petaluma, CA 94952, USA}
\altaffiltext{11}{Department of Physics, Lehigh University, Bethlehem, PA 18015, USA}
\altaffiltext{12}{Department of Astronomy, The University of Tokyo, 7-3-1 Hongo, Bunkyo-ku, Tokyo 113-0033, Japan}
\altaffiltext{13}{Astrobiology Center, NINS, 2-21-1 Osawa, Mitaka, Tokyo 181-8588, Japan}
\altaffiltext{14}{National Astronomical Observatory of Japan, NINS, 2-21-1 Osawa, Mitaka, Tokyo 181-8588, Japan}
\altaffiltext{15}{SOKENDAI (The Graduate University for Advanced Studies), 2-21-1 Osawa, Mitaka, Tokyo 181-8588, Japan}
\altaffiltext{16}{Okayama Astrophysical Observatory, National Astronomical Observatory of Japan, Asakuchi, Okayama 719-0232, Japan}
\altaffiltext{17}{Department of Physical Sciences, Kutztown University, Kutztown, PA 19530, USA}
\altaffiltext{18}{Department of Physics and Astronomy, University of Louisville, Louisville, KY 40292, USA}
\altaffiltext{19}{Department of Physics and Astronomy, Swarthmore College, Swarthmore, PA 19081, USA}
\altaffiltext{20}{Atalaia Group and CROW Observatory, Portalegre, Portugal}
\altaffiltext{21}{Ankara Universitesi Fen Fak. Astronomi ve Uzay Bil Bol. E Blok 205 TR-06100 Tandogan, Ankara TURKEY}
\altaffiltext{22}{Department of Physics, Westminster College, New Wilmington, PA 16172, USA}
\altaffiltext{23}{Department of Physics, Grinnell College, Grinnell, IA 50112, USA}
\altaffiltext{24}{Societ\`a Astronomica Lunae, Castelnuovo Magra 19030, Italy}
\altaffiltext{25}{Astronomy Department, California Institute of Technology, Pasadena, CA, USA}
\altaffiltext{26}{Astronomy Department, University of California, Berkeley, CA, USA}
\altaffiltext{27}{Institut de Recherche sur les Exoplan\`etes, Universit\'e de Montr\'eal, Montre\'eal, QC, Canada}
\altaffiltext{28}{Trottier Fellow}
\altaffiltext{29}{Acton Sky Portal, 3 Concetta Circle, Acton, MA 01720, USA}
\altaffiltext{30}{Department of Astronomy and Astrophysics, The Pennsylvania State University, University Park, PA 16802, USA}
\altaffiltext{31}{Center for Exoplanets and Habitable Worlds, The Pennsylvania State University, University Park, PA 16802, USA}
\altaffiltext{32}{Sagan Fellow}
\altaffiltext{33}{Las Cumbres Observatory Global Telescope Network, Santa Barbara, CA 93117, USA}
\altaffiltext{34}{ICO, Adelaide, South Australia}
\altaffiltext{35}{Department of Physics and Astronomy, Brigham Young University, Provo, UT 84602, USA}
\altaffiltext{36}{Spot Observatory, Nashville, TN 37206, USA}
\altaffiltext{37}{Dipartimento di Fisica “E. R. Caianiello,” Università di Salerno, Via Giovanni Paolo II 132, 84084 Fisciano (SA), Italy}
\altaffiltext{38}{Istituto Internazionale per gli Alti Studi Scientifici (IIASS), Via G. Pellegrino 19, 84019 Vietri sul Mare (SA), Italy}
\altaffiltext{39}{Hazelwood Observatory, Churchill, Victoria, Australia}
\altaffiltext{40}{Perth Exoplanet Survey Telescope, Perth, Australia}
\altaffiltext{41}{George P. and Cynthia Woods Mitchell Institute for Fundamental Physics and Astronomy, Texas A \& M University, College Station, TX 77843, USA}
\altaffiltext{42}{Department of Physics and Astronomy, Texas A \& M University, College Station, TX 77843, USA}
\altaffiltext{43}{Winer Observatory, Sonoita, AZ 85637, USA}
\altaffiltext{44}{NASA Goddard Space Flight Center, Exoplanets and Stellar Astrophysics Laboratory (Code 667), Greenbelt, MD 20771, USA}

\shorttitle{KELT-18\MakeLowercase{b}}

\begin{abstract}
We report the discovery of KELT-18b, a transiting hot Jupiter in a 2.87d orbit around the bright ($V=10.1$), hot, F4V star BD+60 1538 (TYC 3865-1173-1).  We present follow-up photometry, spectroscopy, and adaptive optics imaging that allow a detailed characterization of the system.  Our preferred model fits yield a host stellar temperature of $\teffvaltwo$ K and a mass of $\mstarvaltwo\msun$, situating it as one of only 
a handful of known transiting planets with hosts that are as hot, 
massive,
and bright.
The planet has a mass of $\mpvaltwo\mj$, a radius of $\rpvaltwo\rj$, and a density of $\rhopvaltwo~\densitycgs$, making it one of the most inflated planets known around a hot star.  We argue that KELT-18b's high temperature and low surface gravity, which yield an estimated $\sim 600$ km atmospheric scale height, combined with its hot, bright host make it an excellent candidate for observations aimed at atmospheric characterization.  We also present evidence for a bound stellar companion at a projected separation of $\sim1100$ AU, and speculate that it may have contributed to the strong misalignment we suspect between KELT-18's spin axis and its planet's orbital axis.  The inferior conjunction time is $2457542.524998 \pm 0.000416$ (\bjdtdb) and the orbital period is $2.8717510 \pm 0.0000029$ days.   We encourage Rossiter-McLaughlin measurements in the near future to confirm the suspected spin-orbit misalignment of this system.
\end{abstract}


\section{Introduction}
\label{sec:Intro}
In the 17 years since the detection of the first known transiting exoplanet HD209458 \citep{Charbonneau:2000,Henry:2000}, transit surveys have come of age and refined our understanding of exoplanetary system architectures.  We now know of several thousand transiting planets.  While the \textit{Kepler} space mission \citep{Borucki:2010} was responsible for most of these discoveries, the majority of the $\sim300$ with masses $>0.5\mj$ were discovered by ground-based surveys that are optimized to find giant planets in short-period orbits, now called hot Jupiters.  

One such survey is the Kilodegree Extremely Little Telescope (KELT) project \citep{Pepper:2007, Pepper:2012} which has been carrying out synoptic observations of the sky for more than a decade. Owned and operated by Ohio State, Vanderbilt, and Lehigh Universities, KELT features two 42mm diameter telescopes, one in Arizona (KELT-North) and one in South Africa (KELT-South).  Each has a $26\arcdeg\times26\arcdeg$ field of view and a pixel scale of $23\arcsec$, and together they survey $>70\%$ of the sky with a cadence of 10-20 min and photometric precision of $\sim1\%$.  KELT aims to detect transits of stars in the magnitude range $8\lesssim V\lesssim 11$, filling a niche between the brighter stars generally targeted by radial-velocity surveys and fainter stars measured by other transit surveys.  This is also the range that will be covered in the forthcoming Transiting Exoplanet Survey Satellite TESS \citep{Ricker:2015}, for which KELT is laying the groundwork along with other successful transit searches including the
Hungarian-made Automated Telescope Network (HATNet/HATSouth; \citealt{Bakos:2004,Bakos:2013}), the XO Project \citep{McCullough:2005}, the Wide Angle Search for Planets (SuperWASP; \citealt{Pollacco:2006}, the Trans-Atlantic Exoplanet Search (TrES;\citealt{Alonso:2004}), and the Qatar Exoplanet Survey \citep{Alsubai:2013}.  Newer searches in this category that have already begun science operations include the Multi-site All-Sky CAmeRA\footnote{http://mascara1.strw.leidenuniv.nl}, Everyscope \footnote{http://evryscope.astro.unc.edu}, and the Next-Generation Transit Survey\footnote{http://www.ngtransits.org}.

Even though the number of known transiting hot Jupiters has grown, some regions of their parameter space are still sparsely sampled.  As discussed in \citet{Bieryla:2015}, the KELT survey includes a higher percentage of hot, luminous stars than do transit surveys targeting fainter stars.  Coupled with the fact that surveys like KELT are biased towards finding planets on the large end of the underlying radius distribution \citep{Gaudi:2005} this means that KELT is efficient in detecting strongly irradiated and inflated giant planets.  Their bright hosts make them excellent candidates for follow-up observations and detailed characterization with a range of techniques, and their extremes in host temperature and planet radius make them useful for constraining models of hot Jupiter formation and evolution.

In this paper, we present the discovery of KELT-18b, an inflated hot Jupiter orbiting a hot $V=10.1$ mag F4 star.  KELT-18b joins a still-small collection of very low density, highly inflated planets transiting hot hosts. We describe KELT and follow-up photometry, spectroscopy, and adaptive optics imaging observations (\S\ref{sec:Obs}), and we use them to characterize stellar and planetary parameters (\S\ref{sec:Star},\S\ref{sec:Planet}).  We also report the detection of a faint neighboring star and consider whether it is a bound companion (\S\ref{sec:Neighbor}).  Finally, we situate KELT-18b in the landscape of hot Jupiters and discuss its potential to provide constraints on models of hot Jupiter formation and evolution (\S\ref{sec:Discussion}).

\section{Discovery and Follow-Up Observations}
\label{sec:Obs}

\subsection{Discovery}
\label{sec:Discovery}

\begin{figure}
\centering 
\includegraphics[width=1.1\columnwidth, angle=0]{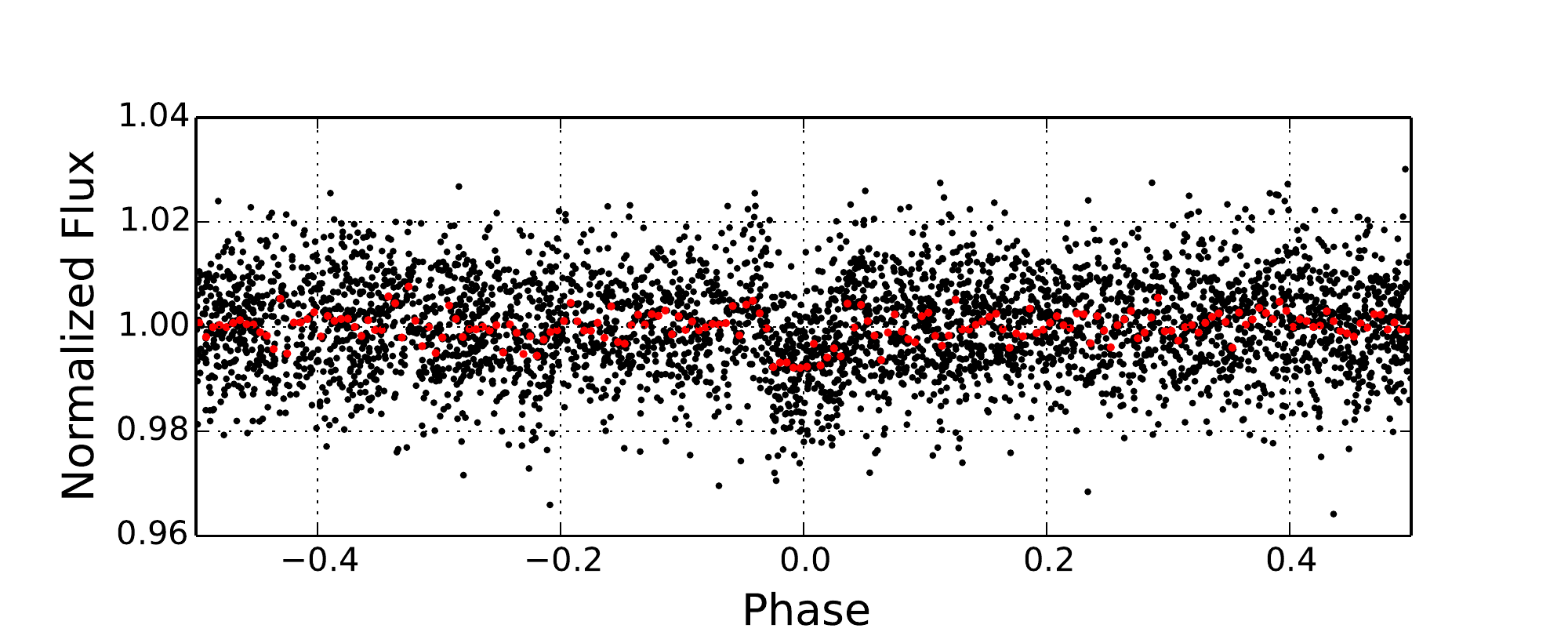}
\caption{\footnotesize Discovery light curve for KELT-18b based on 4162 observation from the KELT-North telescope.  The data have been phase-folded on the preliminary value for the period, 2.8716482d.}
\label{fig:DiscoveryLC}
\end{figure}

The star BD+60 1538 = TYC 3865-1173-1 = KELT-18 was identified as a candidate host star of a transiting planet in KELT-North field 21, a $26\arcdeg\times26\arcdeg$ region centered on $(\alpha,\delta) = (\rm13.39h,+57.0\arcdeg)$. The discovery light curve, shown in Figure \ref{fig:DiscoveryLC}, was based on 4162 observations obtained between 2012 Feb and 2014 Dec.  A Box-Least-Squares \citep{Kovacs:2002} analysis yielded a preliminary period of $\rm 2.8716482~d$, duration of $\rm 4.14~h$, and depth of $\rm 6.8~mmag$. A detailed description of KELT image analysis procedures is given in \citet{Siverd:2012}.  A summary of KELT-18 photometric and kinematic properties is given in Table \ref{tab:LitProps}.

\begin{table}
\footnotesize
\centering
\caption{Collected and Determined KELT-18 properties}
\begin{tabular}{llcc}
  \hline
  \hline
Other identifiers\dotfill & 
        \multicolumn{3}{l}{BD+60 1538} \\
      & \multicolumn{3}{l}{TYC 3865-1173-1}				\\
	  & \multicolumn{3}{l}{2MASS J14260576+5926393}
\\
\hline
Parameter & Description & Value & Ref. \\
\hline
$\alpha_{\rm J2000}$\dotfill	&Right Ascension (RA)\dotfill & $14^h26^m05\fs78$			& 1	\\
$\delta_{\rm J2000}$\dotfill	&Declination (Dec)\dotfill & +59\arcdeg26\arcmin39\farcs24			& 1	\\
\\
$NUV$\dotfill           & GALEX $NUV$ mag.\dotfill & 13.804 $\pm$ 0.004 & 2 \\
$FUV$\dotfill           & GALEX $FUV$ mag.\dotfill & 18.466 $\pm$ 0.056 & 2 \\
\\
$B_{\rm T}$\dotfill			&Tycho $B_{\rm T}$ mag.\dotfill & 10.711 $\pm$ 0.037		& 1	\\
$V_{\rm T}$\dotfill			&Tycho $V_{\rm T}$ mag.\dotfill & 10.214 $\pm$ 0.033		& 1	\\
$B$\dotfill		& APASS Johnson $B$ mag.\dotfill	& 10.534 $\pm$	0.064		& 3	\\
$V$\dotfill		& APASS Johnson $V$ mag.\dotfill	& 10.117 $\pm$	0.022		& 3	\\
$g'$\dotfill		& APASS Sloan $g'$ mag.\dotfill	& 10.595 	 $\pm$	0.106		& 3	\\
$r'$\dotfill		& APASS Sloan $r'$ mag.\dotfill	& 	10.043	$\pm$ 0.016	& 3	\\
$i'$\dotfill		& APASS Sloan $i'$ mag.\dotfill	& 	10.031 $\pm$	0.021	& 3	\\
$z'$\dotfill        & Sloan $z'$ mag.\dotfill & 10.2 $\pm$ 0.1& \S\ref{sec:SED}\\
\\
$J$\dotfill			& 2MASS $J$ mag.\dotfill & 9.454  $\pm$ 0.031		& 4, 5	\\
$H$\dotfill			& 2MASS $H$ mag.\dotfill & 9.272 $\pm$ 0.036	& 4, 5	\\
$K_{\rm S}$\dotfill			& 2MASS $K_{\rm S}$ mag.\dotfill & 9.210 $\pm$ 0.022	& 4, 5	\\
\\
\textit{WISE1}\dotfill		& \textit{WISE1} mag.\dotfill & 9.135 $\pm$ 0.023		& 6, 7	\\
\textit{WISE2}\dotfill		& \textit{WISE2} mag.\dotfill & 9.154 $\pm$ 0.020		& 6, 7 \\
\textit{WISE3}\dotfill		& \textit{WISE3} mag.\dotfill &  9.170 $\pm$ 0.029		& 6, 7	\\
\textit{WISE4}\dotfill		& \textit{WISE4} mag.\dotfill & $\ge 9.085 $		& 6, 7	\\
\\
$\mu_{\alpha}$\dotfill		& Gaia DR1 proper motion\dotfill & $-19.71 \pm$ 1.37 		& 8 \\
                    & \hspace{3pt} in RA (mas yr$^{-1}$)	& & \\
$\mu_{\delta}$\dotfill		& Gaia DR1 proper motion\dotfill 	&  6.09 $\pm$ 1.11 &  8 \\
                    & \hspace{3pt} in DEC (mas yr$^{-1}$) & & \\
\\
$RV$\dotfill & Systemic radial \hspace{9pt}\dotfill  & $-11.6 \pm 0.1$ & \S\ref{sec:Spectra} \\
     & \hspace{3pt} velocity (\kms)  & & \\
$v\sin{i_\star}$\dotfill &  Stellar rotational \hspace{7pt}\dotfill &  12.3 $\pm$ 0.3 & \S\ref{sec:SpecParams} \\
                 & \hspace{3pt} velocity (\kms)  & & \\
Spec. Type\dotfill & Spectral Type\dotfill & F4V & \S\ref{sec:Evol} \\
Age\dotfill & Age (Gyr)\dotfill & $1.9 \pm 0.2$ & \S\ref{sec:Evol} \\
$d$\dotfill & Distance (pc)\dotfill & $311 \pm 14$ & \S\ref{sec:SED} \\
$A_V$\dotfill & Visual extinction (mag) & $0.015\substack{-0.015 \\+0.020}$ & \S\ref{sec:SED} \\
$U^{*}$\dotfill & Space motion (\kms)\dotfill & $-15.9\pm 2.1 $  & \S\ref{sec:UVW} \\
$V$\dotfill       & Space motion (\kms)\dotfill & $-7.8\pm 1.7$ & \S\ref{sec:UVW} \\
$W$\dotfill       & Space motion (\kms)\dotfill & $3.1\pm 1.1$ & \S\ref{sec:UVW} \\
\hline
\hline
\end{tabular}
\begin{flushleft} 
 \footnotesize{ \textbf{\textsc{NOTES:}}
all photometric apertures include the neighbor at $3\farcs4$ (see \S\ref{sec:AO})
$^{*}U$ is positive in the direction of the Galactic Center. 
    References are: $^1$\citet{Hog:2000}, $^2$\citet{Bianch:2011}, $^3$\citet{Henden:2015}, $^4$\citet{Cutri:2003}, $^5$\citet{Skrutskie:2006}, $^6$\citet{Wright:2010}, $^7$\citet{Cutri:2014}, $^8$\citet{Brown:2016} Gaia DR1 http://gea.esac.esa.int/archive/ 
}
\end{flushleft}
\label{tab:LitProps}
\end{table}

\subsection{Photometric Follow-up from KELT-FUN}
\label{sec:Photom}

Once KELT candidates are identified, they are disseminated to a world-wide team of collaborators spanning $\approx60$ institutions known as the KELT-Follow Up Network (KELT-FUN\footnote{For partial lists of KELT-FUN partners with links to individual observatories, see the KELT-North and KELT-South websites at http://www.astronomy.ohio-state.edu/keltnorth/Home.html and 
https://my.vanderbilt.edu/keltsouth/}).  KELT-FUN time-series photometry is used to verify transits, improve ephemerides, and refine transit parameters.  These observations also help to identify false positives such as blends and nearby eclipsing binaries that lie inside the large apertures used with the KELT telescopes' $23\arcsec$ pixels.  KELT-FUN observers plan photometric follow-up observations with the help of custom software tools including a web-based transit prediction calculator based on the TAPIR package \citep{Jensen:2013}.  Observers reduce their own data and generate preliminary light curves that are submitted to the KELT science team.  When an exoplanet is confirmed, the individual follow-up images are collected by the science team and final aperture photometry is carried out using the AstroImageJ package (AIJ; \citealt{Collins:2016, Collins:2013})\footnote{\url{http://www.astro.louisville.edu/software/astroimagej}}. All times are converted to barycentric Julian dates at mid-exposure, \bjdtdb \citep{Eastman:2010}.  

\begin{figure}
  \centering
\includegraphics[width=1\linewidth,clip,height=5.2in]{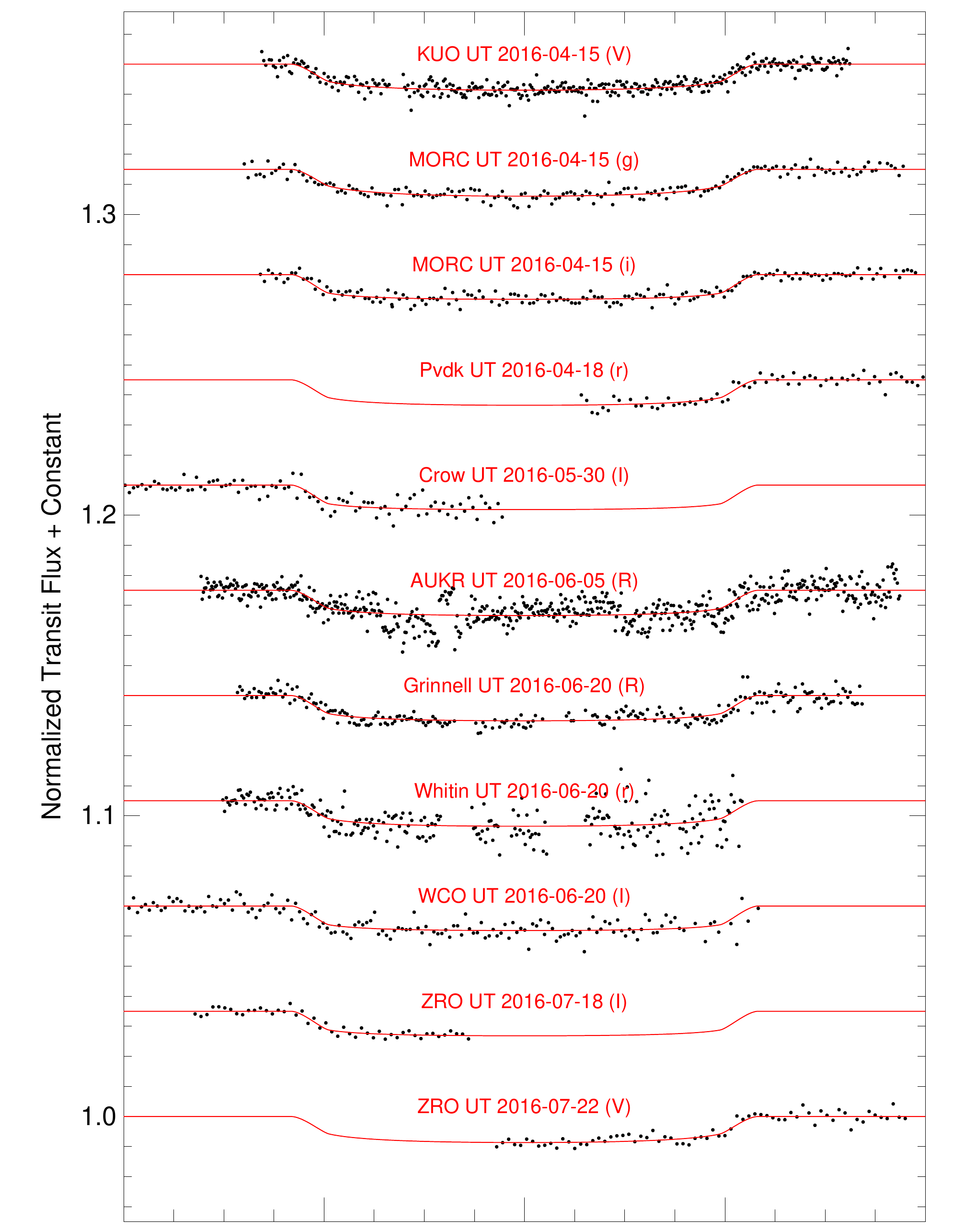}

\vspace{-.1in}
\centering\includegraphics[width=1\linewidth,clip]{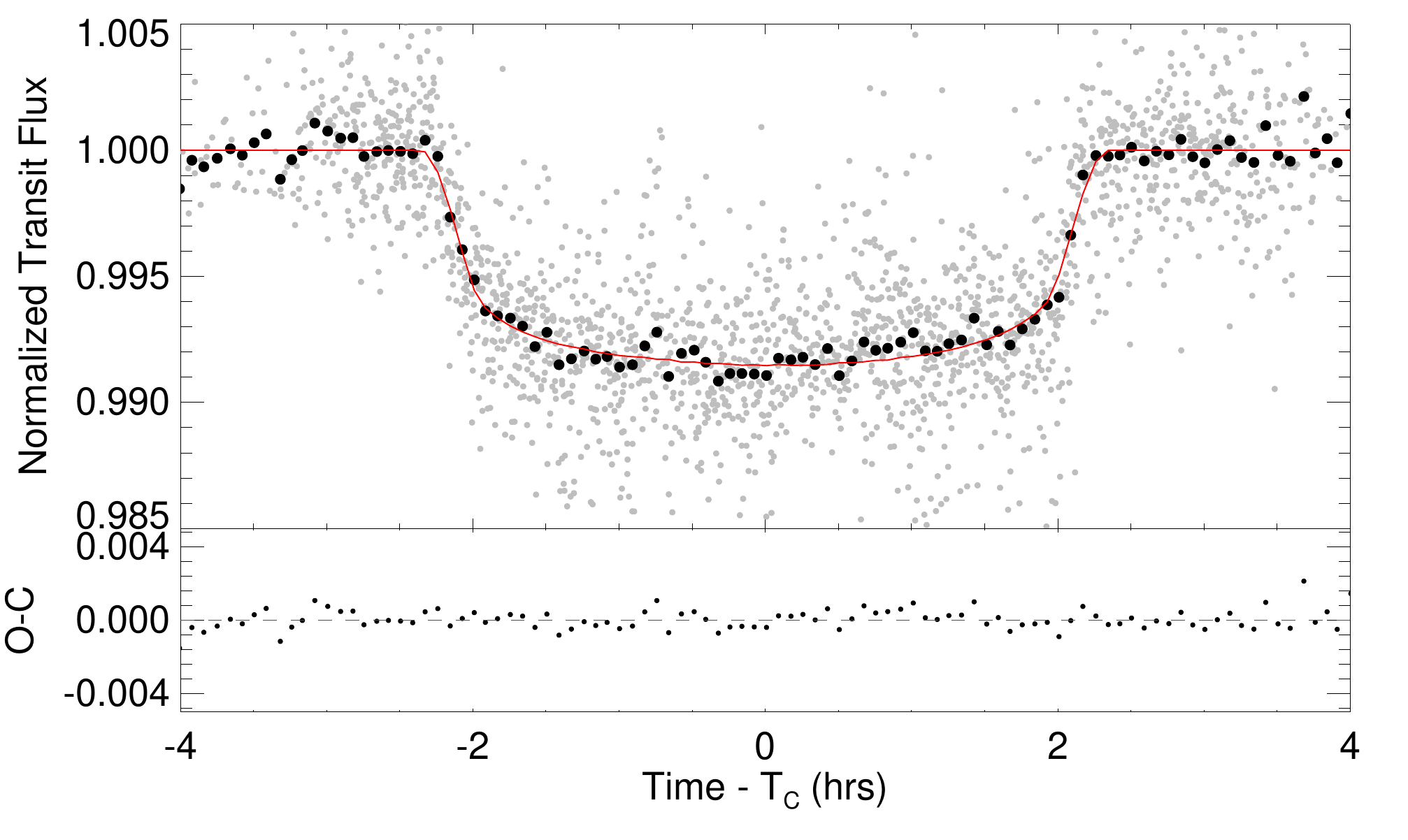}
\caption{\footnotesize KELT-18 light curves.  {\it Top:} The individual transit observations of KELT-18b from KELT-FUN (points) shown with the best fit model from the global fit (\S\ref{sec:GlobalFit}) overplotted (red line).  Note the consistent apparent depths across the various passbands (filters are given in parentheses). {\it Bottom:} Combined KELT-FUN photometry both unbinned (grey points) and binned in 5 minute intervals, along with the residuals (O-C) from the model. Combined data are not used in the analysis but help to illustrate consistency among the light curves.}
\label{fig:All_Lightcurve} 
\end{figure}

The KELT-FUN photometric observations used in our KELT-18 analysis are summarized in Table \ref{tab:Photom}.  We obtained five full transits and six usable partials between UT 2016 Apr 15 and 2016 Jul 21 at nine different observatories:  Ankara University Krieken Observatory (AUKR), Canela's Robotic Observatory (CROW), Grinnell College Grant O. Gale Observatory (Grinnell), Kutztown University Observatory (KUO), the University of Louisville Moore Observatory Ritchey-Chr\'{e}tien (MORC) telescope, Swarthmore College Peter van de Kamp Observatory (PvdK), Westminster College Observatory (WCO), Wellesley College Whitin Observatory (Whitin), and  Roberto Zambelli's Observatory (ZRO).  The individual and combined light curves are shown in Figure \ref{fig:All_Lightcurve}.  

When producing the preliminary individual light curves in AIJ, we fit a transit model to each data set with limb-darkening parameters chosen appropriately for the stellar type.  We use the Bayesian Information Criterion to select the best complement of comparison stars for each dataset, as well as to determine which observed parameters may be systematically affecting the differential photometry.  These ``detrending parameters," shown in Table \ref{tab:Photom}, are then included as free parameters when incorporating the KELT-FUN photometric data sets in the global fits (see \S \ref{sec:GlobalFit} below).  Airmass is always included as a detrending parameter, as even differential photometry may suffer airmass-dependent effects particularly when the comparison stars have different colors from the target.  Other parameters considered include the position of the target on the CCD (``x" and  ``y"), which may induce trends due to residual flatfielding and illumination patterns; the AIJ estimate of the full width at half maximum (FWHM) of the stellar images ( ``FWHM"); the summed counts in the ensemble of comparison stars ( ``total counts"); the sky brightness near the target ( ``sky/pixel"); and a constant offset that can be applied to discontinuous data (denoted  ``meridian flip" as it often results from position shifts that occur when a meridian crossing requires a telescope flip on some equatorial mounts).

We also include in Table \ref{tab:Photom} the sizes of the apertures used for photometry.  KELT-FUN photometric aperture sizes vary from data set to data set because of the differences in plate scale and seeing conditions at this diverse collection of observatories, plus the fact that some observers intentionally defocus to minimize the effects of flatfielding errors.  Optimal photometric apertures are determined individually for each data set.  We include them here to allow an assessment of possible contamination from any neighboring objects.

\begin{table*}
 \footnotesize
 \centering
 \setlength\tabcolsep{1.5pt}
 \caption{Photometric follow-up observations of KELT-18\MakeLowercase{b}}
 \begin{tabular}{lcccccccc}
   \hline
   \hline

Observatory & Location & Aperture & Plate scale& Date      & Filter  & $r^a$ & Exposure & Detrending parameters$^b$  \\
            &          & (m)      & ($\rm \arcsec~pix^{-1}$)& (UT 2016) &        & ($\arcsec$) & Time (s) &  \\
\hline

KUO         & PA       & 0.6      & 0.72       & Apr 15    & V       & 7.9   & 60       & airmass\\
MORC        & KY       & 0.6      & 0.39       & Apr 15    & g$\prime$     & 11.7  & 40       & airmass, sky/pixel, FWHM\\
MORC        & KY       & 0.6      & 0.39       & Apr 15    & i$\prime$     & 11.7  & 50       & airmass, sky/pixel, FWHM\\
PvdK        & PA       & 0.6      & 0.76       & Apr 18    & r$\prime$     & 9.2   & 45       & airmass, x\\
CROW        & Portugal & 0.30     & 0.82       & May 30    & I$_c$   & 8.3   &120       & airmass\\
AUKR      & Turkey   & 1.0      & 0.78       & Jun 05    & R       & 9.3   & 40       & airmass, sky/pixel\\
Grinnell    & IA       & 0.6      & 0.37       & Jun 20    & R       & 7.4   & 80       & airmass, x, y, total counts, FWHM\\
Whitin      & MA       & 0.6      & 0.58       & Jun 20    & r$\prime$     & 8.1   & 43       & airmass, x, meridian flip\\
WCO         & PA       & 0.35     & 1.45       & Jun 20    & I       & 8.7   & 30       & airmass, FWHM\\
ZRO         & Italy    & 0.4      & 0.52       & Jul 18    & I$_c$   & 8.8   &200       & airmass\\
ZRO         & Italy    & 0.4      & 0.52       & Jul 21    & V       & 7.7   &200       & airmass\\
\hline
\hline

\end{tabular}
\begin{flushleft}
  \footnotesize{ \textbf{\textsc{NOTES:}} $^a$Radius of photometric aperture.  $^b$Photometric parameters allowed to vary in global fits and described in the text.}
\end{flushleft}
\label{tab:Photom}
\end{table*}

\subsection{Spectroscopic Follow-up}
\label{sec:Spectra}

We obtained high resolution spectra of KELT-18 to measure radial velocities (RVs), to make sure that the candidate is not a double-lined spectroscopic binary, to help rule out a stellar companion by checking that the RVs yield a spectroscopic orbit in agreement with the photometric ephemeris, and to determine the stellar spectral parameters.

We used the Tillinghast Reflector Echelle Spectrograph (TRES\footnote{http://tdc-www.harvard.edu/instruments/tres}; \citealt{Szentgyorgyi:2007, Furesz:2008}) on the 1.5 m telescope at the Fred Lawrence Whipple Observatory (FLWO) on Mt. Hopkins, Arizona, as well as the Levy high-resolution optical spectrograph on the 2.4m Automated Planet Finder (APF\footnote{http://www.ucolick.org/public/telescopes/apf.htm}; \citealt{Vogt:2014}) at Lick Observatory on Mt. Hamilton, California. We have a total of 17 TRES and 11 APF observations taken between UT 2016 Apr 15 and Jun 24, summarized in Table \ref{tab:Spectra} and included in the global fits.  

The TRES spectra have resolution $R \sim 44,000$ and were obtained using a $2\farcs3$ fiber.  They were reduced, extracted, and RV-analyzed as described by \citet{Buchave:2010}.  We derive {\it relative} radial velocities (RVs) by cross-correlating each observed spectrum order by order against the observed spectrum with the highest S/N in the wavelength range 4300-5660 \AA. The observation used as the reference is assigned an RV of 0 \kms \hspace{2 pt} by definition.  We find {\it absolute} radial velocities by cross correlating the Mg b line region against a synthetic template spectrum generated using the \citet{Kurucz:1992} stellar atmosphere models.  To find the systemic RV, we took a weighted average of the individual TRES absolute velocities after correcting each for the phase-dependent orbital velocity based on the orbital fit. The systemic RV was then adjusted to the International Astronomical Union (IAU) Radial Velocity Standard Star system \citep{Stefanik:1999} via a correction of -0.6 \kms\  primarily to correct for the gravitational redshift, which is not included in the library of synthetic template spectra. We find the absolute velocity of the KELT-18 system to be -11.6 $\pm$ 0.1 \kms, where the uncertainty is an estimate of the residual systematic errors in the transfer to the IAU system.

The APF spectra have resolution $R \sim 100,000$ and were obtained using a $1\arcsec\times3\arcsec$ slit.  They were reduced, extracted and RV-analyzed as described in \citet{Fulton:2015}. The star was observed through a cell of gaseous iodine which imprints a dense forest of molecular absorption lines onto the stellar spectrum to serve as both a wavelength and PSF fiducial.  Because this star is relatively faint for the APF with its extremely high spectral resolution and modest aperture size, it was impractical to collect the high S/N iodine-free template needed in the RV forward modeling process \citep{Butler:1996}. We instead collected a single observation of KELT-18 using the HIRES instrument on Keck 1 \citep{Vogt:1994} to be used as the iodine-free template in the RV extraction. We exposed for 248 s using the 0.86\arcsec x 3.5\arcsec\ Decker B5 slit for an effective spectral resolution of R$\sim$50,000 and S/N$\sim$100. We deconvolved this stellar template with the instrumental point-spread-function derived from observations of rapidly rotating B stars observed through the iodine cell.

None of the spectroscopic measurements were made during a transit, so we cannot yet carry out a Rossiter-McLaughlin (RM, \citet{Rossiter:1924,McLaughlin:1924}) or Doppler tomographic (DT) analysis to determine the relative alignment of the projected stellar spin axis and planetary orbital axis.  

\begin{table}
\centering
 \caption{Relative RVs and bisectors for KELT-18\MakeLowercase{b}}
 \label{tab:Spectra}
 \begin{tabular}{lrrrrl}
    \hline
    \hline
    \multicolumn{1}{l}{\bjdtdb} & \multicolumn{1}{c}{RV} 	& \multicolumn{1}{c}{$\sigma_{\rm RV}$} 	& \multicolumn{1}{c}{Bisector} &  \multicolumn{1}{c}{$\sigma_{\rm Bisector}$}	& Source \\
    & \multicolumn{1}{c}{($\rm m~s^{-1}$)} &\multicolumn{1}{c}{($\rm m~s^{-1}$)} &
	    \multicolumn{1}{c}{($\rm m~s^{-1}$)} &\multicolumn{1}{c}{($\rm m~s^{-1}$)} \\
    \hline                                                                                       
2457502.83243 & -228.5 & 42.0  &   19.7  &  24.7  & TRES\\
2457512.82572 &   13.2 & 59.2  &   35.3  &  40.6  & TRES\\
2457514.83486 & -131.3 & 33.6  &  -38.0  &  22.5  & TRES\\
2457532.76871 &  -45.8 & 53.8  &    8.8  &  31.2  & TRES\\
2457533.78374 &   26.3 & 42.4  &   10.8  &  20.3  & TRES\\
2457534.68911 & -121.0 & 50.0  &   27.0  &  22.9  & TRES\\
2457535.92184 &  173.0 & 49.6  &  -27.1  &  34.6  & TRES\\
2457536.92732 &  -92.0 & 26.7  &  -29.3  &  26.3  & TRES\\
2457537.71324 & -142.5 & 40.4  &   32.0  &  23.0  & TRES\\
2457538.74047 &   66.2 & 26.9  &  -15.6  &  14.8  & TRES\\
2457539.78838 & -129.2 & 32.0  &    4.2  &  17.3  & TRES\\
2457550.67430 &   55.8 & 36.2  &   -5.8  &  18.7  & TRES\\
2457551.73053 & -196.5 & 26.4  &    2.8  &  17.2  & TRES\\
2457552.68605 &  -52.4 & 23.0  &    9.5  &  14.1  & TRES\\
2457553.68429 &   15.3 & 20.4  &  -11.8  &  14.1  & TRES\\
2457554.76974 & -167.4 & 23.7  &  -10.9  &  13.5  & TRES\\
2457555.66513 &    0.0 & 20.4  &  -11.7  &  15.4  & TRES\\
2457496.86075 & -117.2 & 21.2  &  -77.4  & 132.6  &  APF\\
2457498.78813 &  219.4 & 27.4  &    7.4  &  40.9  &  APF\\
2457507.69757 &  106.7 & 27.3  &  239.0  &  76.1  &  APF\\
2457509.68051 &   -7.0 & 28.3  &  305.9  & 155.4  &  APF\\
2457521.01283 &   94.4 & 20.0  & -185.8  &  77.4  &  APF\\
2457524.73222 &   90.9 & 23.6  &  -50.3  &  63.4  &  APF\\
2457531.72812 &  -77.0 & 36.4  & -268.5  & 169.1  &  APF\\
2457533.73676 &  -83.4 & 25.4  &  183.6  &  90.2  &  APF\\
2457537.78009 & -155.6 & 20.1  &  177.3  & 120.5  &  APF\\
2457541.69419 &   54.0 & 25.5  &  -21.7  & 118.3  &  APF\\
2457543.69475 & -121.8 & 22.6  & -133.1  & 137.1  &  APF\\
    \hline
    \hline
 \end{tabular}
 \begin{flushleft}
  \footnotesize{ \textbf{\textsc{NOTES:}}  The TRES RVs zeropoint is arbitrarily set to the last TRES value; AFP RVS have an arbitrary zeropoint that is within $\sim25\ \rm m\ s^{-1}$ from zero.  Both can be fit as free parameters in subsequent analyses.
  }
\end{flushleft}
\end{table}

\begin{figure*}
\includegraphics[width=1\linewidth]{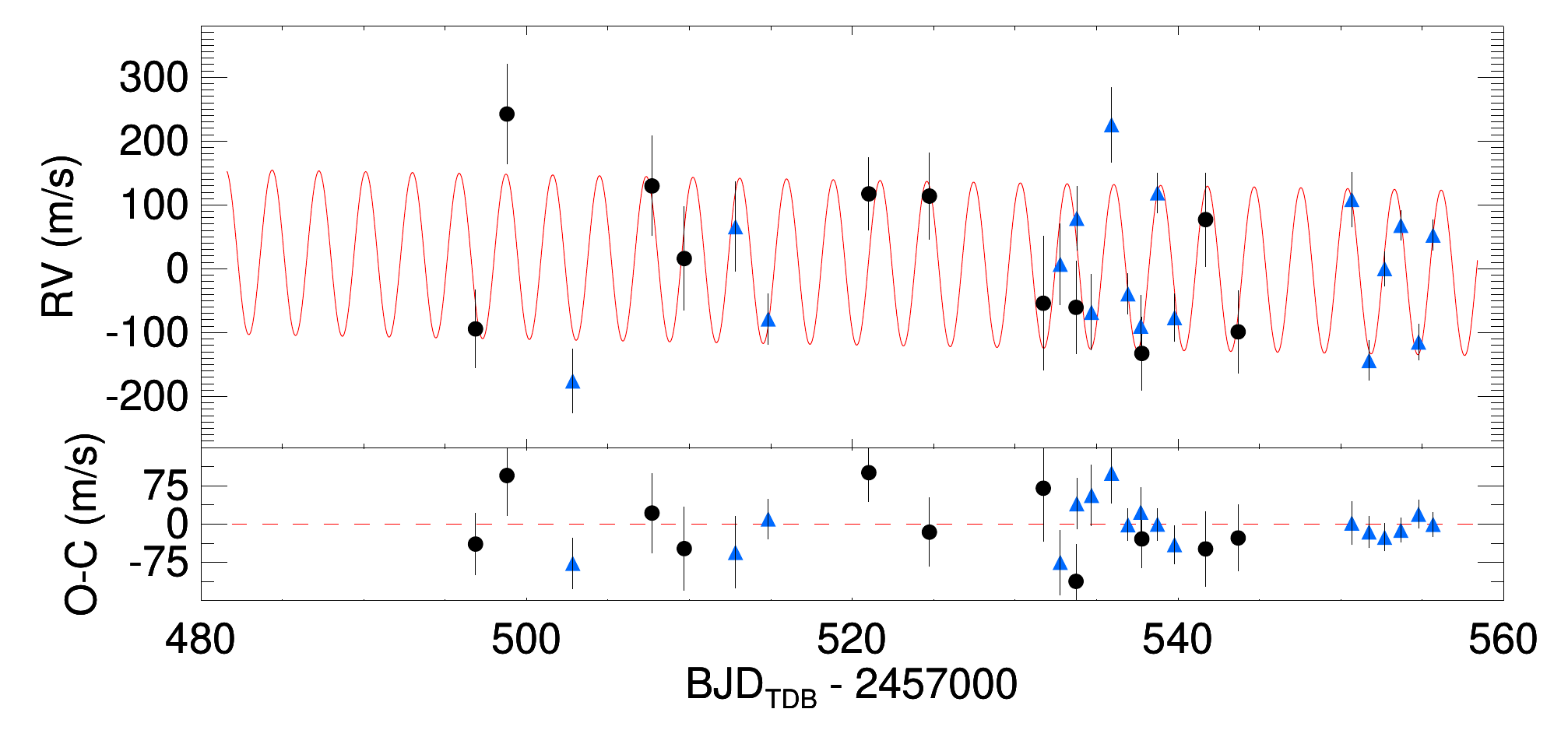}
   \vspace{-.3in}
\includegraphics[width=1\linewidth]{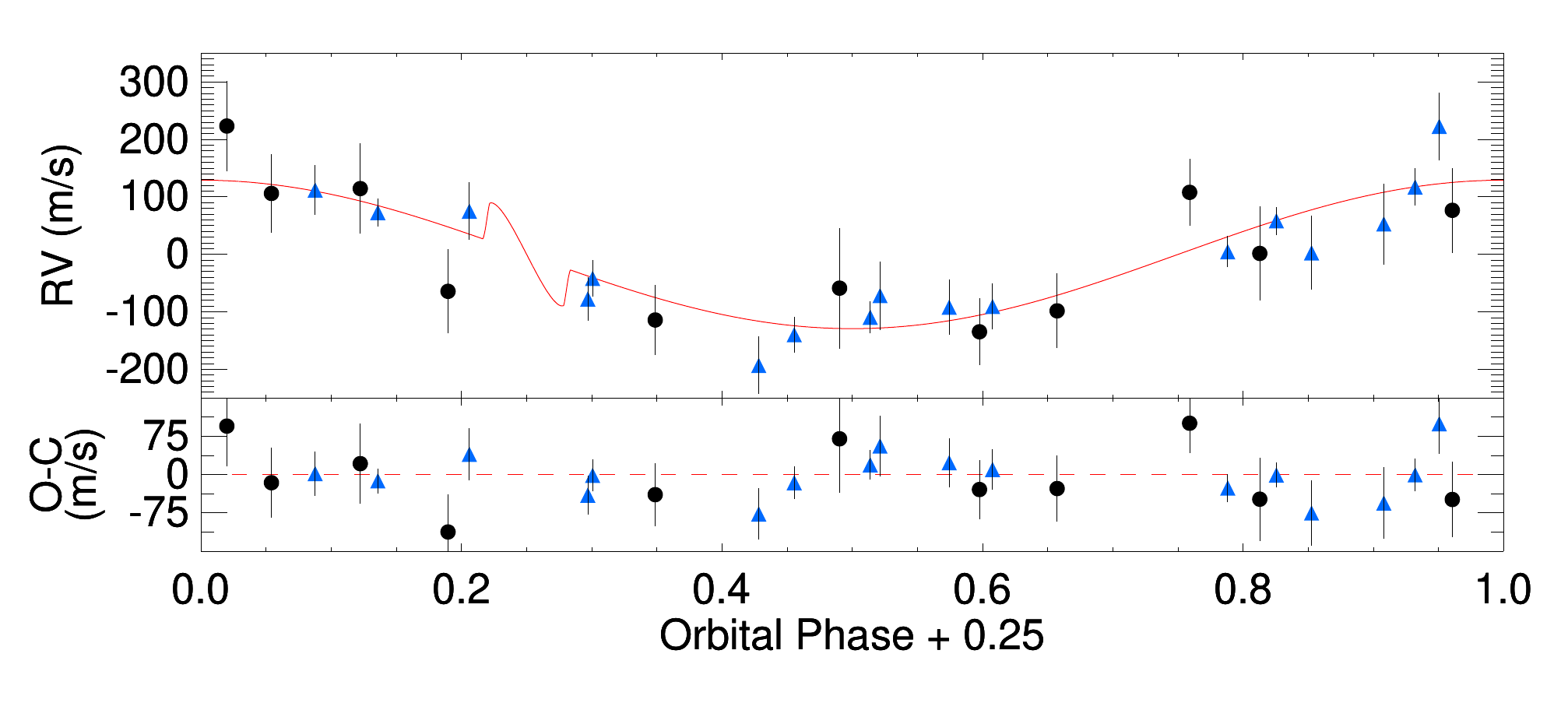}
   \vspace{.1in}
\caption{\footnotesize KELT-18 radial velocities.  (Top) Relative radial velocity measurements by TRES (blue triangles) and APF (black dots) along with the best fit orbit plus a slow downward drift (red). The residuals are shown below. Our global fits in  \S\ref{sec:GlobalFit} indicate that the slope of the drift is not statistically significant.  (Bottom) Relative RV measurements of KELT-18 phase-folded to the period from our global model. The feature centered at phase 0.25 represents the RM prediction for in-transit radial velocity measurements.  We have adopted the YY circular fit from \S\ref{sec:GlobalFit} throughout.
}
\label{fig:RVs} 
\end{figure*}

\subsection{High Contrast AO Imaging with Subaru IRCS}
\label{sec:AO}

To check for the presence of other nearby stars that might contaminate the target and affect the interpretation, we obtained KELT-18 observations on 2016 Jun 27 with the Infrared Camera and Spectrograph (IRCS; \citealt {Kobayashi:2000}) along with the 188-element adaptive optics system AO188 \citep{Hayano:2010} on the Subaru 8.2m telescope. We employed the high-resolution mode of IRCS, which gives a pixel scale of $\rm 20.6\ mas~pix^{-1}$ and a field of view of $ 21\farcs1 \times 21\farcs1$. The target star itself was used as a natural guide star. We observed the target with the $K^\prime$-band filter at five dithering positions, each with the exposure time of 10 s. We took four sets of dithered images without a neutral density (ND) filter to search for faint companion candidates and one set with a 1\% ND filter to avoid saturation of the target. The airmass at the time of observation was 1.3. The FWHM of the target with AO was 0\farcs12.

\begin{figure}
\includegraphics[width=1\linewidth]{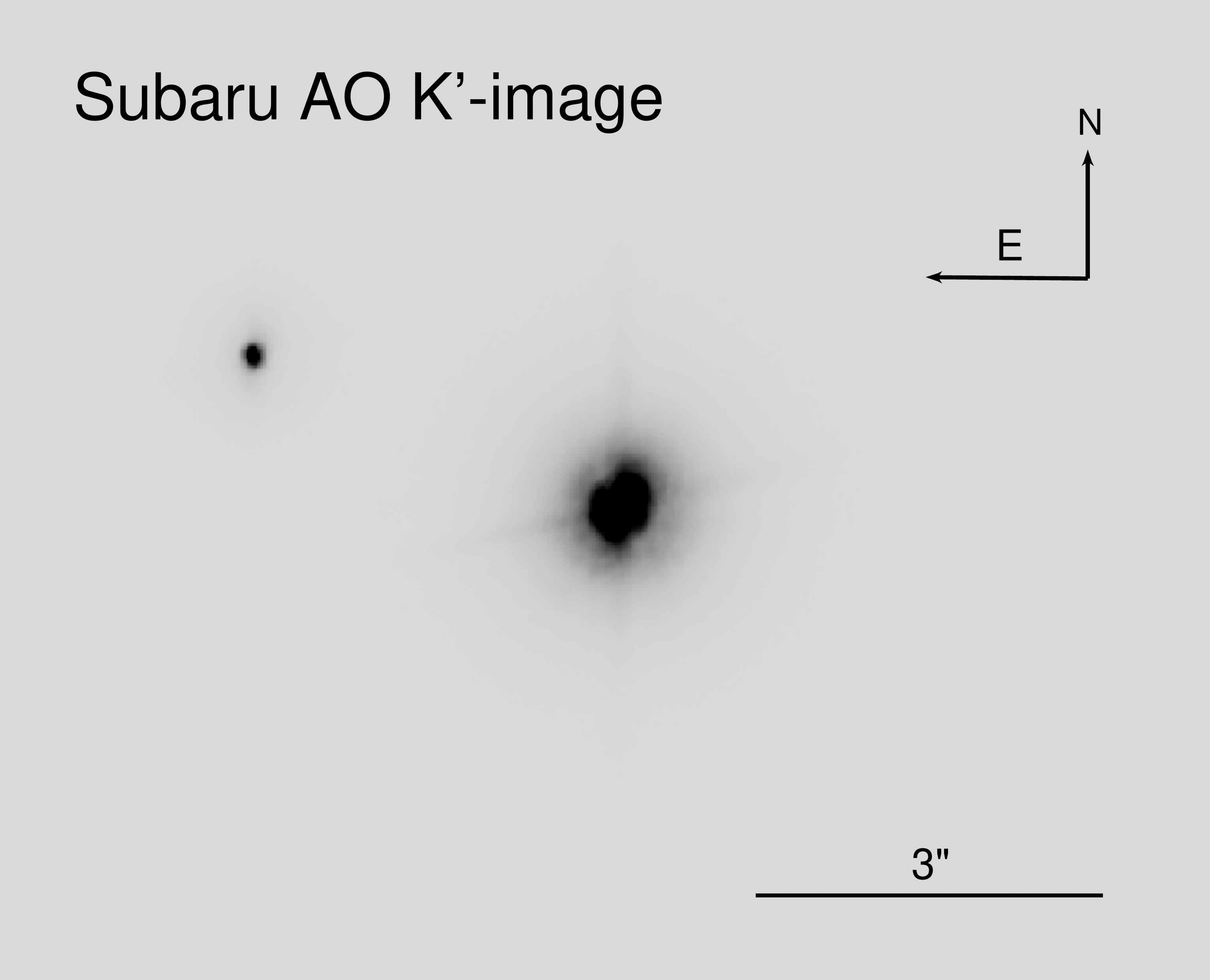}
\caption{\footnotesize Subaru AO $K'_S$ image of KELT-18, clearly showing the neighbor at separation $3\farcs43 \pm 0\farcs01$ and PA=67\arcdeg.  The greyscale stretch was chosen to highlight the detection limits; no significance should be attributed to KELT-18's apparent shape, which is the combined result of saturation and an asymmetrical PSF.  Flux ratios were measured using an unsaturated image.}

\label{fig:AOim}
\end{figure}

\begin{figure}
\includegraphics[width=1\linewidth]{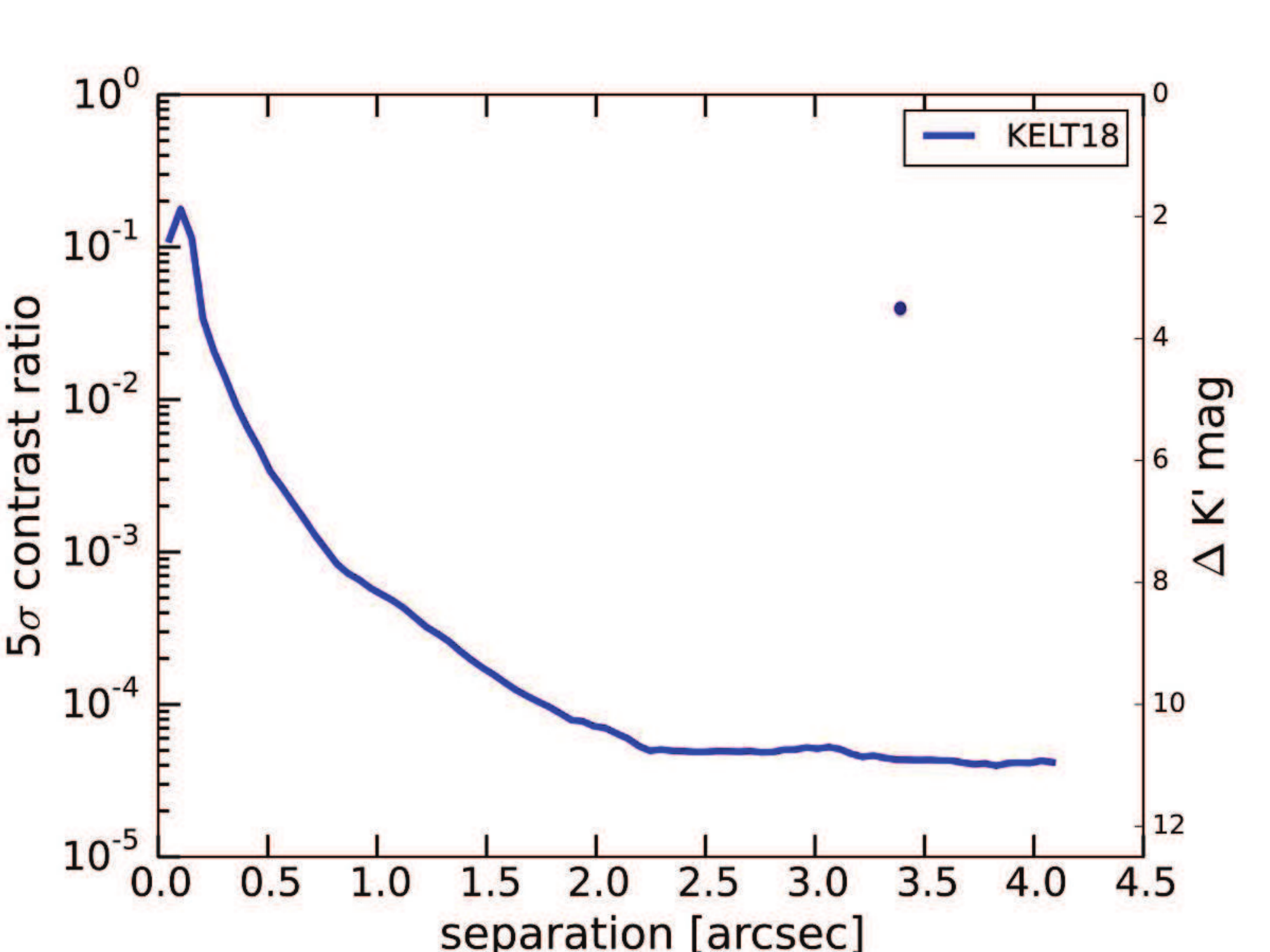}
\caption{\footnotesize Subaru AO $5\sigma$ contrast curve, showing the limiting flux ratio and magnitude for the detection of a point source as a function of separation from the target. The KELT-18 neighbor is cleanly detected, as shown by the dot (larger than the uncertainties for clarity).  We see no other neighbors.}
\label{fig:AOContrast}
\end{figure}

The observed images were dark subtracted and flatfielded in a standard manner. The reduced images were then aligned, sky-level subtracted, and median combined. The combined image from the set taken without the ND filter and the $5\sigma$ contrast curve are shown in Figures \ref{fig:AOim} and \ref{fig:AOContrast}. A faint neighbor was easily detected at a separation of $3\farcs43 \pm 0\farcs01$ at a position angle PA=67\arcdeg\ (east of north).  (Note: the shape of the core of KELT-18 in the image is an artifact of saturation plus an asymmetrical PSF that can also be seen on the fainter neighbor when viewed at a higher stretch).

Using the ND image in which KELT-18 is unsaturated, we measure a magnitude difference between the neighbor and KELT-18 of $\Delta K^\prime = 3.6 \pm 0.2$, corresponding to a flux ratio of $28\pm5$. Taking the combined magnitude of the two stars to be $K_S=9.210\pm0.022$ (Table \ref{tab:LitProps}), we calculate the neighbor's apparent magnitude to be $K_S=12.9\pm0.2$.   We see no other companions above the contrast threshold.  The proximity of the neighbor means that all of the photometry listed in Table \ref{tab:LitProps} as well as our ground-based follow-up photometry will include the light of the neighbor, which we correct for as described in \S\ref{sec:SED} below.  The spectroscopic apertures are small enough to be uncontaminated.

\section{Host Star Characterization}
\label{sec:Star}

\subsection{Spectral Analysis}
\label{sec:SpecParams}
We obtain initial estimates of some of the KELT-18 physical properties from the TRES spectra using the Spectral Parameter Classification (SPC) procedure of \citet{Buchave:2012} (see also \citet{Torres:2012} for a comparison of SPC with other procedures). Running SPC with no parameters fixed, taking the error-weighted mean value for each stellar parameter, and adopting the mean error for each parameter we get an effective temperature of $\rm \teff=6634\pm120\ K$, a surface gravity of $\loggstar=3.93\pm0.19\ \rm \gravitycgs$, a metallicity of $ \meh = -0.09\pm0.13$, and a projected equatorial rotation speed of $\vsinistar=12.3\pm0.3\ \kms$ (the last may be an overestimate as SPC does not explicitly account for macroturbulence). These values are used as starting points and/or priors to help constrain the initial global fits as described in \S\ref{sec:GlobalFit}, which in turn generate refined values for these parameters.

 We also obtained estimates for the stellar parameters based on our Keck spectrum using the SpecMatch procedure \citep{Petigura:2015}:
 $\rm \teff=6538\pm60\ K$, 
 $\loggstar=4.13\pm0.07\ \rm \gravitycgs$, 
 $ \feh = -0.12\pm0.04$, and 
 $\vsinistar=10\pm1\ \kms$.
 However, because the temperature puts it outside the range over which SpecMatch is calibrated ($\rm teff<6250 K$), we did not use these values in the analysis.

\subsection{SED Analysis}
\label{sec:SED}
We use KELT-18's spectral energy distribution (SED) to determine its distance and reddening.  Table \ref{tab:LitProps} lists the near-UV to mid-IR fluxes that we have compiled from the literature for KELT-18.  However, all of these fluxes were measured through photometric apertures that contain the faint neighbor (see \S\ref{sec:AO}), so before we carry out an SED analysis we need to account for the neighbor's contributions to the broadband fluxes.  We know the KELT-18-to-neighbor flux ratio in $K^\prime$ from the AO observations (\S\ref{sec:AO}).  The neighbor can also be seen in the SDSS {\it z-}, {\it i-}, and possibly {\it r-}band images, though any measurement is complicated by the fact that KELT-18 itself is saturated in all three bands.  

We were able to estimate the KELT-18-to-neighbor flux ratio in the $z$-band using a combination of SDSS images, our own follow-up images, and APASS photometry. For this we carried out PSF fits using the C program {\it imfitfits} provided by Brian McLeod and described in \citet{Lehar:2000}. \textit{Imfitfits} makes a model by convolving theoretical point sources with an observed PSF, and then varying any combination of the parameters defining the background level and point source positions and magnitudes to minimize the sum of the squares of the residuals over all the pixels. We modeled KELT-18 on the SDSS {\it z} image as two point sources whose relative positions were constrained by the AO images, using another star in the field as an empirical PSF and masking out the saturated KELT-18 center during fitting.  From the best-fit parameters of the two-star fit, we generated a model containing only KELT-18 and subtracted it from the original image to obtain an image of the neighbor only. From that we measured the neighbor's flux in a small aperture and converted to magnitudes using the SDSS zero point from the image header and an aperture correction empirically determined from other stars on the frame.  For the neighbor we estimate $z=14.6\pm0.1$.  Because there is no APASS magnitude in {\it z}, we determined the KELT-18 magnitude in {\it z} using one of our follow-up images (PvdK).  We performed aperture photometry of KELT-18 (including the neighbor) and four unsaturated, SDSS-cataloged stars on the same image and find $z=10.2\pm0.1$ mag.  (Here the uncertainty represents the scatter among values derived from the ensemble of comparison stars; there are few that are both faint enough to be unsaturated in SDSS and bright enough to be visible in the short follow-up images).  From our estimates we calculate a magnitude difference of $\Delta z = 4.4 \pm 0.2$ corresponding to a flux ratio of $\sim60$ in {\it z}.

\begin{figure}
\vspace{0.1in}
\includegraphics[width=1\linewidth]{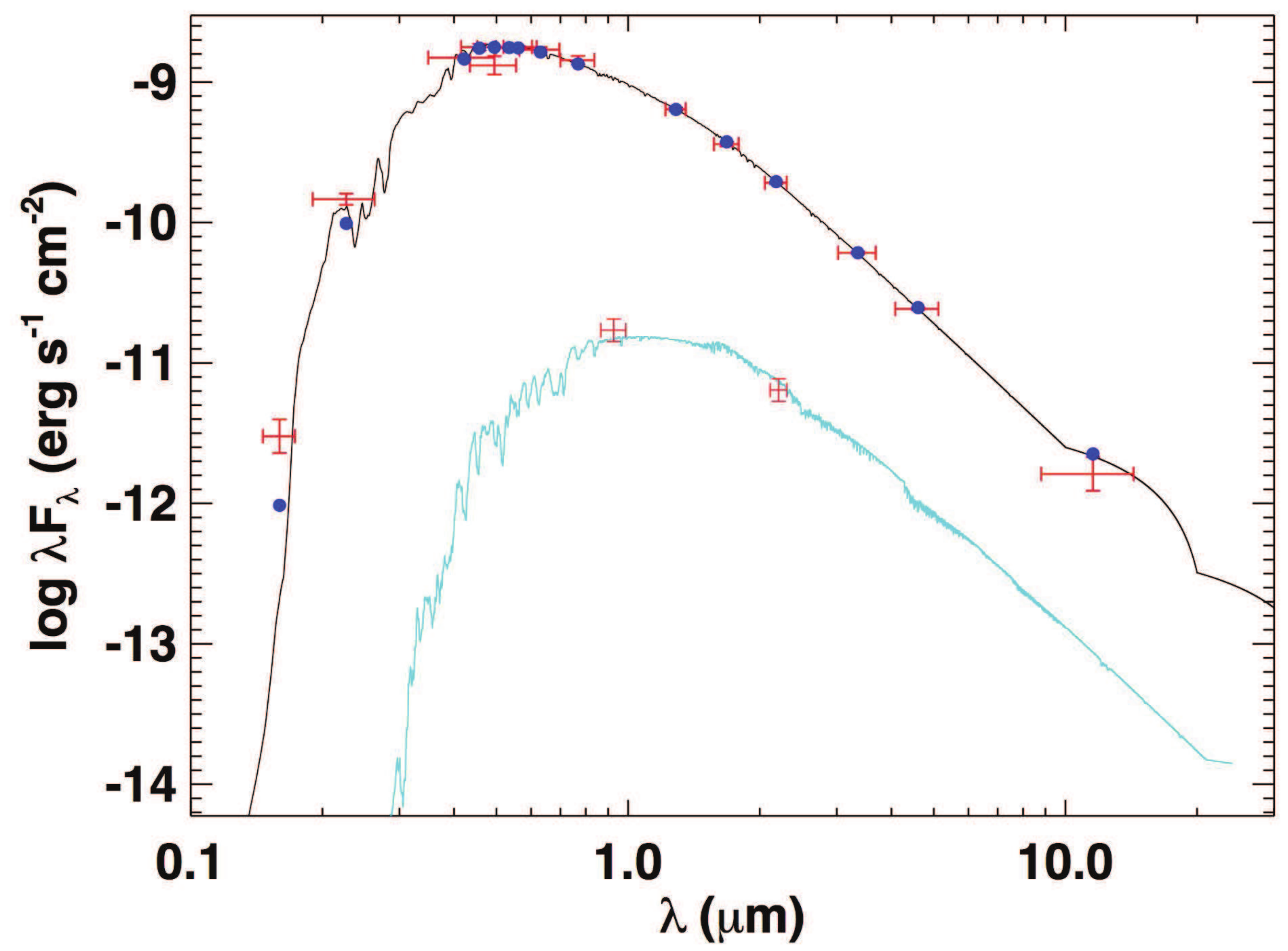}
\caption{\footnotesize SED fits to KELT-18 and its faint neighbor. Red crosses show observed values with vertical errorbars representing $1\sigma$ measurement uncertainty and horizontal errorbars representing the width of each bandpass.  Blue points give the model fluxes in the observed bandpasses.  Solid lines show the model fits.  The faint, redder neighbor contributes $\lesssim1\%$ to the combined bolometric flux.}
\label{fig:SED}
\end{figure}

Armed with the flux ratios in $K^\prime$ and {\it z}, we fit the Table \ref{tab:LitProps} fluxes using the \citet{Kurucz:1992} model atmospheres.  We fix the KELT-18 values of
$\teff=6670\pm 120~\rm K$, $\loggstar= 4.056_{-0.014}^{+0.011}$, and $ [Fe/H]=0.07\pm 0.13$ 
from the initial global fit (\S\ref{sec:GlobalFit}) assuming a circular orbit and using the Yonsei-Yale stellar evolution models (\citet{Demarque:2004}; hereafter ``YY").  The results are shown in Figure \ref{fig:SED}.  We find the 
neighbor's temperature to be $\teff\sim3900~\rm K$ with an overall contribution of $<$ 1\% to the SED.  For the individual filters in our KELT-FUN photometry the neighbor's contributions to the fluxes computed from the SEDs are {$F_{\rm neighbor} / F_{\rm KELT-18}$} = 0.4\% ($B$), 0.5\% ($V$), 0.8\% ($R$), 0.8\% ($I$), 0.6\% ($g$), 0.8\% ($r$), 0.9\% ($i$), and 1\% ($z$).  (We note that the $z$ value is only marginally consistent with our measured value, which translates to a flux ratio of $1.7\pm 0.3\%$, but the difference is not enough to affect the analysis.)

Adjusting for the neighbor's contribution in each passband, we find KELT-18's visual extinction to be $ A_V=0.015_{-0.015}^{+0.020}$ and its bolometric flux to be $ F_{bol} = 2.14\times~\fluxcgs$ with an error of $\sim3\%$.  
From this bolometric flux and the values of the KELT-18 luminosity, radius, and temperature from the final global fit (\S\ref{sec:GlobalFit} and Table \ref{tab:KELT-18b_global_fit_properties}), we compute the KELT-18 distance to be $\rm 311\pm14\ pc$.  This is consistent with the Gaia first data release\footnote{Gaia DR1 http://gea.esac.esa.int/archive/} value of $\rm  331\pm56\ pc$ \citep{Brown:2016} but with a much smaller uncertainty.  We note though that our uncertainty includes only the formal uncertainty on each parameter and does not include any component that might arise from the use of different stellar models.

\subsection{Evolutionary Analysis}
\label{sec:Evol}

\begin{figure}
\includegraphics[width=1\linewidth]{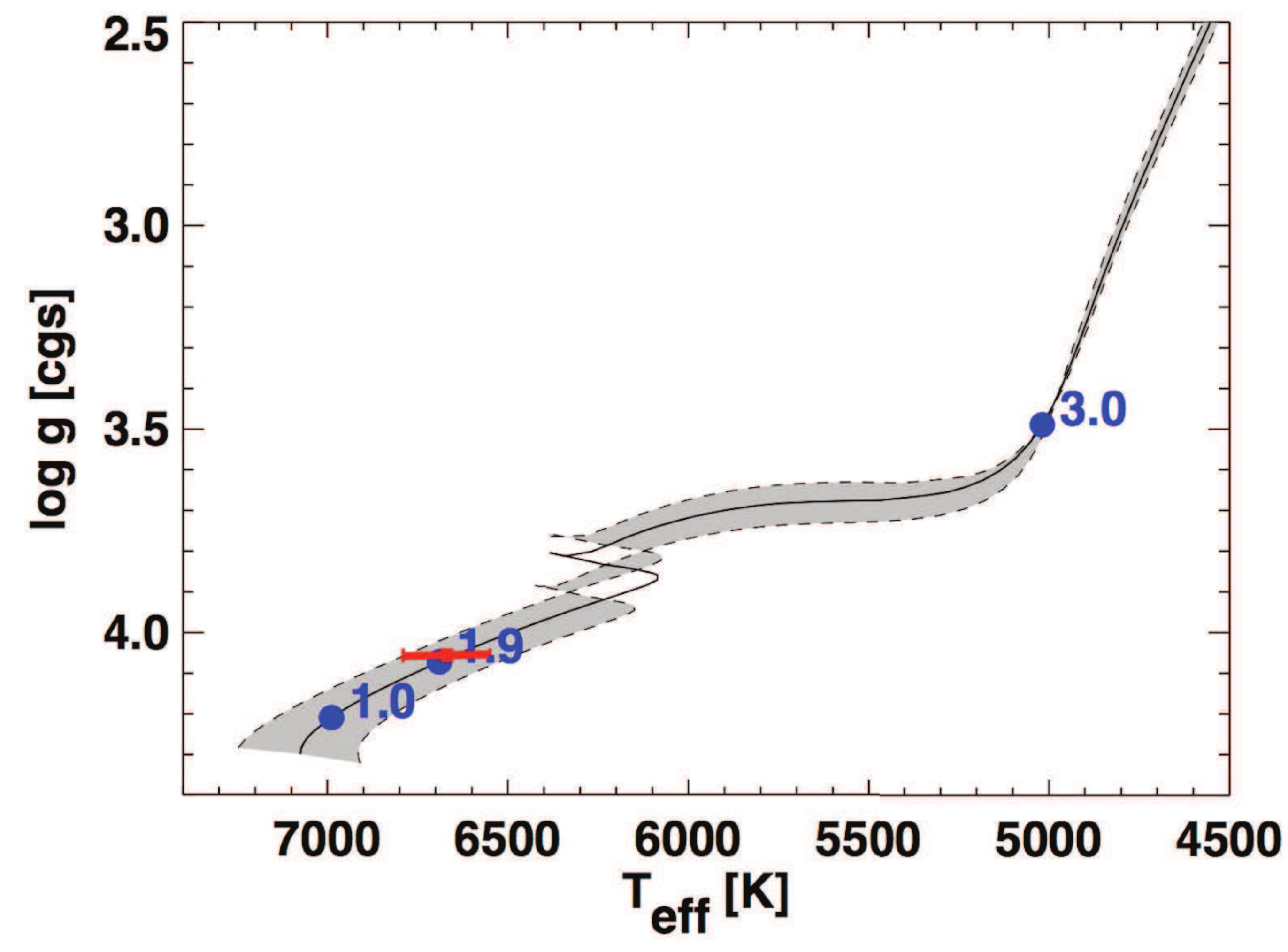}
\caption{\footnotesize Age determination for the KELT-18 host.  We fit the $M_*$ and \feh~ from the global analysis to the YY evolutionary models. The shaded region represents the 1$\sigma$ regime for the models, and the blue markers give the ages in Gyr along the best-fit track.  We find an age of 1.7-2.1 Gyr for KELT-18, which is shown in red along with the $1\sigma$ uncertainties on \teff~ and \loggstar~ from the global fits. 
}
\label{fig:Evol}
\end{figure}

We use the stellar parameters from the YY circular case in \S\ref{sec:GlobalFit} and Table \ref{tab:KELT-18b_global_fit_properties} to determine the evolutionary state of KELT-18.  Comparing $\teff=6670\ \rm K$ against the tabulation of dwarf stars in \citet{Pecaut:2013}, we find KELT-18 to have a spectral type of F4.  To estimate its age, we follow the procedure specified in \citet{Siverd:2012} and subsequent KELT discovery papers to match the stellar parameters to YY evolutionary tracks.  We select the evolutionary tracks based on the $M_*$ and \feh~from our initial fits in \S\ref{sec:GlobalFit} and compare the predicted \teff~ and \loggstar~ to our measured values. The tracks are shown in Figure \ref{fig:Evol}.   We find an age of 1.9 $\pm$ 0.2 Gyr, where the uncertainty includes only the propagation of the uncertainties in the stellar parameters from the global fit, and does not include systematic or calibration uncertainties of the YY model itself.  We conclude that KELT-18 is a main sequence F4V star that is about two-thirds of the way through its main sequence lifetime.

\subsection{UVW Space Motion}
\label{sec:UVW}
We have computed the three-dimensional space motion of KELT-18 in an effort to situate it kinematically within a Galactic context. We assume the Table \ref{tab:LitProps} values for the systemic velocity ($-11.6 \pm 0.1$ \kms), distance ($311 \pm 14$ pc), and proper motions (($\mu_{\alpha}$, $\mu_{\delta}$)  = (-19.71 $\pm$ 1.37, 6.09 $\pm$ 1.11) $\rm mas~yr^{-1}$). We adopt the local standard of rest values from \citet{Coskunoglu:2011} and compute the space motions to be ($U$, $V$, $W$) = ($-15.9 \pm 2.1$, $-7.8 \pm 1.7$, $3.1 \pm 1.1$) \kms. Comparison with the distributions in \citet{Bensby:2003} yields a 99.4\% probability that KELT-18 belongs to the Galactic thin disk population. In addition, KELT-18's essentially Solar metallicity and small V velocity are consistent with the young inferred age of $1.9 \pm 0.2$ Gyr (\S\ref{sec:Evol}).

\section{Planet Characterization}
\label{sec:Planet}

\subsection{EXOFAST Global Fit}
\label{sec:GlobalFit}

To determine the physical and observable properties of the KELT-18 system, we conduct global fits as in previous KELT discovery papers.  The technique is explained in detail in \citet{Siverd:2012}; here we provide an overview and describe how the method is applied in the specific case of KELT-18.  We use a modified version of the \cite{Eastman:2013} EXOFAST code, which is an IDL-based fitting tool that runs simultaneous Markov Chain Monte Carlo analyses to determine the posterior probability distribution of each system parameter.  To constrain the host star mass $ M_*$ and radius $ R_*$, EXOFAST can use either the YY stellar evolution models or the empirical relations of \citet{Torres:2010} (hereafter  ``Torres relations"). 

The fitting process is iterative.  For the initial KELT-18 fits, we use the YY models with eccentricity held at zero.  Data inputs to EXOFAST include the relative RV's (\S\ref{sec:Spectra}) and the follow-up time-series photometry along with applicable detrending parameters (\S\ref{sec:Photom}).  Starting values include the orbital period and ephemeris determined from the KELT data, plus the spectroscopically-determined stellar \teff, \feh, \loggstar, and \vsinistar~ (\S\ref{sec:SpecParams}; note we use our \meh\ as a starting point for \feh). Because the light curves are expected to give better constraints on the stellar density  through fitted shapes of the primary transits \citep{Seager:2003,Mortier:2014}, we allow \loggstar~ to vary unconstrained while the other parameters are applied with prior penalties.  For the stellar radius, we include a Gaussian prior calculated from the Stefan-Boltzmann law using the Gaia-determined distance and bolometric flux (\S\ref{sec:SED}) along
with the spectroscopic \teff.  
These initial fits allow us to determine refined values for \teff, \feh, and especially \loggstar, that inform the SED fits (\S\ref{sec:SED}). 

For the next round of fits, we run each of the YY and Torres models for both a circular case and the case where eccentricity is free to vary.  Here we include the corrections to the photometry for extinction and contamination determined from the SED fits (\S\ref{sec:SED}).  We verified that for KELT-18, the neighbor's contribution was so small that the round 2 stellar parameters \teff, \feh, and \loggstar~ from the circular YY fit were unchanged from those in the initial fit; thus we do not need to repeat the SED analysis.  With the results from these fits we carry out a Transit Timing Variation (TTV) analysis (see \S\ref{sec:TTVs}) that yields refined values for the orbital period and time of inferior conjunction.  We run the fits a final time including these as priors.

The results are shown in Tables \ref{tab:KELT-18b_global_fit_properties} and \ref{tab:KELT-18b_global_fit_properties_p2}.  We find that the four fits are consistent with each other to  within $1\sigma$.  The non-circular models result in an eccentricity \ensuremath{0.06_{-0.04}^{+0.07}}, consistent with circular.  Thus, for the analysis in this paper we adopt the YY circular fit.  We find KELT-18b to have radius $R_{P}=\rpvaltwo\rj$, mass $M_{P}=\mpvaltwo\mj$,  and density $\rho_{P}=\rhopvaltwo\ \densitycgs$.  Our fits indicate an RV slope \dotgammavaltwo $\rm\ m\ s^{-1}\ day^{-1}$,  which is consistent with zero.

\begin{table*}
 \scriptsize
\centering
\setlength\tabcolsep{1.5pt}
\caption{Median values and 68\% confidence intervals for the physical and orbital parameters of the KELT-18 system}
  \label{tab:KELT-18b_global_fit_properties}
  \begin{tabular}{lccccc}
  \hline
  \hline
   Parameter & Units & Value &\textbf{Adopted Value} & Value & Value \\
   & & (YY eccentric) & \textbf{(YY circular; $e$=0 fixed)} & (Torres eccentric) &  (Torres circular; $e$=0 fixed) \\
Stellar Parameters & & & & &\\
                                ~~~$M_{*}$\dotfill &                              Mass (\msun)\dotfill &                                      \mstarvalone &                                      \mstarvaltwo &                                    \mstarvalthree &                                     \mstarvalfour\\
                                ~~~$R_{*}$\dotfill &                            Radius (\rsun)\dotfill &                                      \rstarvalone &                                      \rstarvaltwo &                                    \rstarvalthree &                                     \rstarvalfour\\
                                ~~~$L_{*}$\dotfill &                        Luminosity (\lsun)\dotfill &                                      \lstarvalone &                                      \lstarvaltwo &                                    \lstarvalthree &                                     \lstarvalfour\\
                               ~~~$\rho_*$\dotfill &                             Density (\densitycgs)\dotfill &                                        \rhovalone &                                        \rhovaltwo &                                      \rhovalthree &                                       \rhovalfour\\
                            ~~~$\log{g_*}$\dotfill &                     Surface gravity (\gravitycgs)\dotfill &                                   \loggstarvalone &                                   \loggstarvaltwo &                                 \loggstarvalthree &                                  \loggstarvalfour\\
                                ~~~$\teff$\dotfill &                 Effective temperature (K)\dotfill &                                       \teffvalone &                                       \teffvaltwo &                                     \teffvalthree &                                      \teffvalfour\\
                                 ~~~$\feh$\dotfill &                               Metallicity\dotfill &                                        \fehvalone &                                        \fehvaltwo &                                      \fehvalthree &                                       \fehvalfour\\
 \hline
 Planet Parameters & & & & & \\
                                    ~~~$e$\dotfill &                              Eccentricity\dotfill &                                          \evalone &                                          \evaltwo &                                        \evalthree &                                         \evalfour\\
                             ~~~$\omega_*$\dotfill &          Argument of periastron (degrees)\dotfill &                                      \omegavalone &                                      \omegavaltwo &                                    \omegavalthree &                                     \omegavalfour\\
                                    ~~~$P$\dotfill &                             Period (days)\dotfill &                                          \pvalone &                                          \pvaltwo &                                        \pvalthree &                                         \pvalfour\\
                                    ~~~$a$\dotfill &                      Semi-major axis (AU)\dotfill &                                          \avalone &                                          \avaltwo &                                        \avalthree &                                         \avalfour\\
                                ~~~$M_{P}$\dotfill &                                Mass (\mj)\dotfill &                                         \mpvalone &                                         \mpvaltwo &                                       \mpvalthree &                                        \mpvalfour\\
                                ~~~$R_{P}$\dotfill &                              Radius (\rj)\dotfill &                                         \rpvalone &                                         \rpvaltwo &                                       \rpvalthree &                                        \rpvalfour\\
                             ~~~$\rho_{P}$\dotfill &                             Density (\densitycgs)\dotfill &                                       \rhopvalone &                                       \rhopvaltwo &                                     \rhopvalthree &                                      \rhopvalfour\\
                          ~~~$\log{g_{P}}$\dotfill &                           Surface gravity (\gravitycgs)\dotfill &                                      \loggpvalone &                                      \loggpvaltwo &                                    \loggpvalthree &                                     \loggpvalfour\\
                               ~~~$T_{eq}$\dotfill &               Equilibrium temperature (K)\dotfill &                                        \teqvalone &                                        \teqvaltwo &                                      \teqvalthree &                                       \teqvalfour\\
                               ~~~$\Theta$\dotfill &                           Safronov number\dotfill &                                      \thetavalone &                                      \thetavaltwo &                                    \thetavalthree &                                     \thetavalfour\\
                                ~~~$\fave$\dotfill &                  Incident flux (\fluxcgs)\dotfill &                                       \favevalone &                                       \favevaltwo &                                     \favevalthree &                                      \favevalfour\\
 \hline
 RV Parameters & & & & & \\
                                  ~~~$T_C$\dotfill &    Time of inferior conjunction (\bjdtdb)\dotfill &                                         \tcvalone &                                         \tcvaltwo &                                       \tcvalthree &                                        \tcvalfour\\
                                ~~~$T_{P}$\dotfill &              Time of periastron (\bjdtdb)\dotfill &                                         \tpvalone &                                         \tpvaltwo &                                       \tpvalthree &                                        \tpvalfour\\
                                    ~~~$K$\dotfill &                   RV semi-amplitude (m/s)\dotfill &                                          \kvalone &                                          \kvaltwo &                                        \kvalthree &                                         \kvalfour\\
                           ~~~$M_P\sin{i}$\dotfill &                        Minimum mass (\mj)\dotfill &                                     \mpsinivalone &                                     \mpsinivaltwo &                                   \mpsinivalthree &                                    \mpsinivalfour\\
                          ~~~$M_{P}/M_{*}$\dotfill &                                Mass ratio\dotfill &                                    \mpmstarvalone &                                    \mpmstarvaltwo &                                  \mpmstarvalthree &                                   \mpmstarvalfour\\
                                    ~~~$u$\dotfill &                  RM linear limb darkening\dotfill &                                          \uvalone &                                          \uvaltwo &                                        \uvalthree &                                         \uvalfour\\
                         ~~~$\gamma_{APF}$\dotfill &                                       m/s\dotfill &                                   \gammaapfvalone &                                   \gammaapfvaltwo &                                 \gammaapfvalthree &                                  \gammaapfvalfour\\
                        ~~~$\gamma_{TRES}$\dotfill &                                       m/s\dotfill &                                  \gammatresvalone &                                  \gammatresvaltwo &                                \gammatresvalthree &                                 \gammatresvalfour\\
                         ~~~$\dot{\gamma}$\dotfill &                        RV slope (m/s/day)\dotfill &                                   \dotgammavalone &                                   \dotgammavaltwo &                                 \dotgammavalthree &                                  \dotgammavalfour\\
                               ~~~$\ecosw$\dotfill &                                          \dotfill &                                      \ecoswvalone &                                      \ecoswvaltwo &                                    \ecoswvalthree &                                     \ecoswvalfour\\
                               ~~~$\esinw$\dotfill &                                          \dotfill &                                      \esinwvalone &                                      \esinwvaltwo &                                    \esinwvalthree &                                     \esinwvalfour\\
                             ~~~$f(m1,m2)$\dotfill &                       Mass function (\mj)\dotfill &                                  \fmonemtwovalone &                                  \fmonemtwovaltwo &                                \fmonemtwovalthree &                                 \fmonemtwovalfour\\
 \hline
 \end{tabular}
\begin{flushleft}
  \end{flushleft}
\end{table*}

\begin{table*}
 \scriptsize
\centering
\setlength\tabcolsep{1.5pt}
\caption{Median values and 68\% confidence interval for the physical and orbital parameters of the KELT-18 system}
  \label{tab:KELT-18b_global_fit_properties_p2}
  \begin{tabular}{lccccc}
  \hline
  \hline
   Parameter & Units & Value &\textbf{Adopted Value} & Value & Value \\
   & & (YY eccentric) & \textbf{(YY circular; $e$=0 fixed)} & (Torres eccentric) &  (Torres circular; $e$=0 fixed) \\
 Primary Transit & & & & & \\
\hline
                          ~~~$R_{P}/R_{*}$\dotfill &     Radius of the planet in stellar radii\dotfill &                                    \rprstarvalone &                                    \rprstarvaltwo &                                  \rprstarvalthree &                                   \rprstarvalfour\\
                                ~~~$a/R_*$\dotfill &          Semi-major axis in stellar radii\dotfill &                                         \arvalone &                                         \arvaltwo &                                       \arvalthree &                                        \arvalfour\\
                                    ~~~$i$\dotfill &                     Inclination (degrees)\dotfill &                                          \ivalone &                                          \ivaltwo &                                        \ivalthree &                                         \ivalfour\\
                                    ~~~$b$\dotfill &                          Impact parameter\dotfill &                                          \bvalone &                                          \bvaltwo &                                        \bvalthree &                                         \bvalfour\\
                               ~~~$\delta$\dotfill &                             Transit depth\dotfill &                                      \deltavalone &                                      \deltavaltwo &                                    \deltavalthree &                                     \deltavalfour\\
                             ~~~$T_{FWHM}$\dotfill &                      FWHM duration (days)\dotfill &                                      \tfwhmvalone &                                      \tfwhmvaltwo &                                    \tfwhmvalthree &                                     \tfwhmvalfour\\
                                 ~~~$\tau$\dotfill &            Ingress/egress duration (days)\dotfill &                                        \tauvalone &                                        \tauvaltwo &                                      \tauvalthree &                                       \tauvalfour\\
                               ~~~$T_{14}$\dotfill &                     Total duration (days)\dotfill &                                   \tonefourvalone &                                   \tonefourvaltwo &                                 \tonefourvalthree &                                  \tonefourvalfour\\
                                ~~~$P_{T}$\dotfill &  A priori non-grazing transit probability\dotfill &                                         \ptvalone &                                         \ptvaltwo &                                       \ptvalthree &                                        \ptvalfour\\
                              ~~~$P_{T,G}$\dotfill &              A priori transit probability\dotfill &                                        \ptgvalone &                                        \ptgvaltwo &                                      \ptgvalthree &                                       \ptgvalfour\\
                              ~~~$T_{C,0}$\dotfill &                Mid-transit time (\bjdtdb)\dotfill &                                     \tczerovalone &                                     \tczerovaltwo &                                   \tczerovalthree &                                    \tczerovalfour\\
                              ~~~$T_{C,1}$\dotfill &                Mid-transit time (\bjdtdb)\dotfill &                                      \tconevalone &                                      \tconevaltwo &                                    \tconevalthree &                                     \tconevalfour\\
                              ~~~$T_{C,2}$\dotfill &                Mid-transit time (\bjdtdb)\dotfill &                                      \tctwovalone &                                      \tctwovaltwo &                                    \tctwovalthree &                                     \tctwovalfour\\
                              ~~~$T_{C,3}$\dotfill &                Mid-transit time (\bjdtdb)\dotfill &                                    \tcthreevalone &                                    \tcthreevaltwo &                                  \tcthreevalthree &                                   \tcthreevalfour\\
                              ~~~$T_{C,4}$\dotfill &                Mid-transit time (\bjdtdb)\dotfill &                                     \tcfourvalone &                                     \tcfourvaltwo &                                   \tcfourvalthree &                                    \tcfourvalfour\\
                              ~~~$T_{C,5}$\dotfill &                Mid-transit time (\bjdtdb)\dotfill &                                     \tcfivevalone &                                     \tcfivevaltwo &                                   \tcfivevalthree &                                    \tcfivevalfour\\
                              ~~~$T_{C,6}$\dotfill &                Mid-transit time (\bjdtdb)\dotfill &                                      \tcsixvalone &                                      \tcsixvaltwo &                                    \tcsixvalthree &                                     \tcsixvalfour\\
                              ~~~$T_{C,7}$\dotfill &                Mid-transit time (\bjdtdb)\dotfill &                                    \tcsevenvalone &                                    \tcsevenvaltwo &                                  \tcsevenvalthree &                                   \tcsevenvalfour\\
                              ~~~$T_{C,8}$\dotfill &                Mid-transit time (\bjdtdb)\dotfill &                                    \tceightvalone &                                    \tceightvaltwo &                                  \tceightvalthree &                                   \tceightvalfour\\
                              ~~~$T_{C,9}$\dotfill &                Mid-transit time (\bjdtdb)\dotfill &                                     \tcninevalone &                                     \tcninevaltwo &                                   \tcninevalthree &                                    \tcninevalfour\\
                             ~~~$T_{C,10}$\dotfill &                Mid-transit time (\bjdtdb)\dotfill &                                  \tconezerovalone &                                  \tconezerovaltwo &                                \tconezerovalthree &                                 \tconezerovalfour\\
                               ~~~$u_{1I}$\dotfill &                     Linear Limb-darkening\dotfill &                                      \uoneivalone &                                      \uoneivaltwo &                                    \uoneivalthree &                                     \uoneivalfour\\
                               ~~~$u_{2I}$\dotfill &                  Quadratic Limb-darkening\dotfill &                                      \utwoivalone &                                      \utwoivaltwo &                                    \utwoivalthree &                                     \utwoivalfour\\
                               ~~~$u_{1R}$\dotfill &                     Linear Limb-darkening\dotfill &                                      \uonervalone &                                      \uonervaltwo &                                    \uonervalthree &                                     \uonervalfour\\
                               ~~~$u_{2R}$\dotfill &                  Quadratic Limb-darkening\dotfill &                                      \utworvalone &                                      \utworvaltwo &                                    \utworvalthree &                                     \utworvalfour\\
                          ~~~$u_{1Sloang}$\dotfill &                     Linear Limb-darkening\dotfill &                                 \uonesloangvalone &                                 \uonesloangvaltwo &                               \uonesloangvalthree &                                \uonesloangvalfour\\
                          ~~~$u_{2Sloang}$\dotfill &                  Quadratic Limb-darkening\dotfill &                                 \utwosloangvalone &                                 \utwosloangvaltwo &                               \utwosloangvalthree &                                \utwosloangvalfour\\
                          ~~~$u_{1Sloani}$\dotfill &                     Linear Limb-darkening\dotfill &                                 \uonesloanivalone &                                 \uonesloanivaltwo &                               \uonesloanivalthree &                                \uonesloanivalfour\\
                          ~~~$u_{2Sloani}$\dotfill &                  Quadratic Limb-darkening\dotfill &                                 \utwosloanivalone &                                 \utwosloanivaltwo &                               \utwosloanivalthree &                                \utwosloanivalfour\\
                          ~~~$u_{1Sloanr}$\dotfill &                     Linear Limb-darkening\dotfill &                                 \uonesloanrvalone &                                 \uonesloanrvaltwo &                               \uonesloanrvalthree &                                \uonesloanrvalfour\\
                          ~~~$u_{2Sloanr}$\dotfill &                  Quadratic Limb-darkening\dotfill &                                 \utwosloanrvalone &                                 \utwosloanrvaltwo &                               \utwosloanrvalthree &                                \utwosloanrvalfour\\
                               ~~~$u_{1V}$\dotfill &                     Linear Limb-darkening\dotfill &                                      \uonevvalone &                                      \uonevvaltwo &                                    \uonevvalthree &                                     \uonevvalfour\\
                               ~~~$u_{2V}$\dotfill &                  Quadratic Limb-darkening\dotfill &                                      \utwovvalone &                                      \utwovvaltwo &                                    \utwovvalthree &                                     \utwovvalfour\\
Secondary Eclipse & & & & & \\
                                ~~~$T_{S}$\dotfill &                 Time of eclipse (\bjdtdb)\dotfill &                                         \tsvalone &                                         \tsvaltwo &                                       \tsvalthree &                                        \tsvalfour\\
                                ~~~$b_{S}$\dotfill &                          Impact parameter\dotfill &                                         \bsvalone &                                         \bsvaltwo &                                       \bsvalthree &                                        \bsvalfour\\
                           ~~~$T_{S,FWHM}$\dotfill &                      FWHM duration (days)\dotfill &                                     \tsfwhmvalone &                                     \tsfwhmvaltwo &                                   \tsfwhmvalthree &                                    \tsfwhmvalfour\\
                               ~~~$\tau_S$\dotfill &            Ingress/egress duration (days)\dotfill &                                       \tausvalone &                                       \tausvaltwo &                                     \tausvalthree &                                      \tausvalfour\\
                             ~~~$T_{S,14}$\dotfill &                     Total duration (days)\dotfill &                                  \tsonefourvalone &                                  \tsonefourvaltwo &                                \tsonefourvalthree &                                 \tsonefourvalfour\\
                                ~~~$P_{S}$\dotfill &  A priori non-grazing eclipse probability\dotfill &                                         \psvalone &                                         \psvaltwo &                                       \psvalthree &                                        \psvalfour\\
                              ~~~$P_{S,G}$\dotfill &              A priori eclipse probability\dotfill &                                        \psgvalone &                                        \psgvaltwo &                                      \psgvalthree &                                       \psgvalfour\\
     \hline
 \hline
\end{tabular}
\begin{flushleft}
  \footnotesize \textbf{NOTES:} The $T_C$ values are the times of inferior conjunction derived from the individual follow-up light curves        
  \end{flushleft}
\end{table*}

\subsection{Transit Timing Variation Analysis}
\label{sec:TTVs}
We obtain an independent ephemeris by performing a linear fit to all of the follow-up photometry-determined transit center times from our global fit. 
Our analysis gives an inferior conjunction time of $2457542.524998 \pm 0.000416$ (\bjdtdb) and orbital period of $2.8717510 \pm 0.0000029$ days with a $\chi^2$ of 10.69 and 9 degrees of freedom.  We feed these values back into EXOFAST as priors for our final global fits (\S\ref{sec:GlobalFit}).

\begin{figure}[!ht]
\includegraphics[width=1\linewidth]{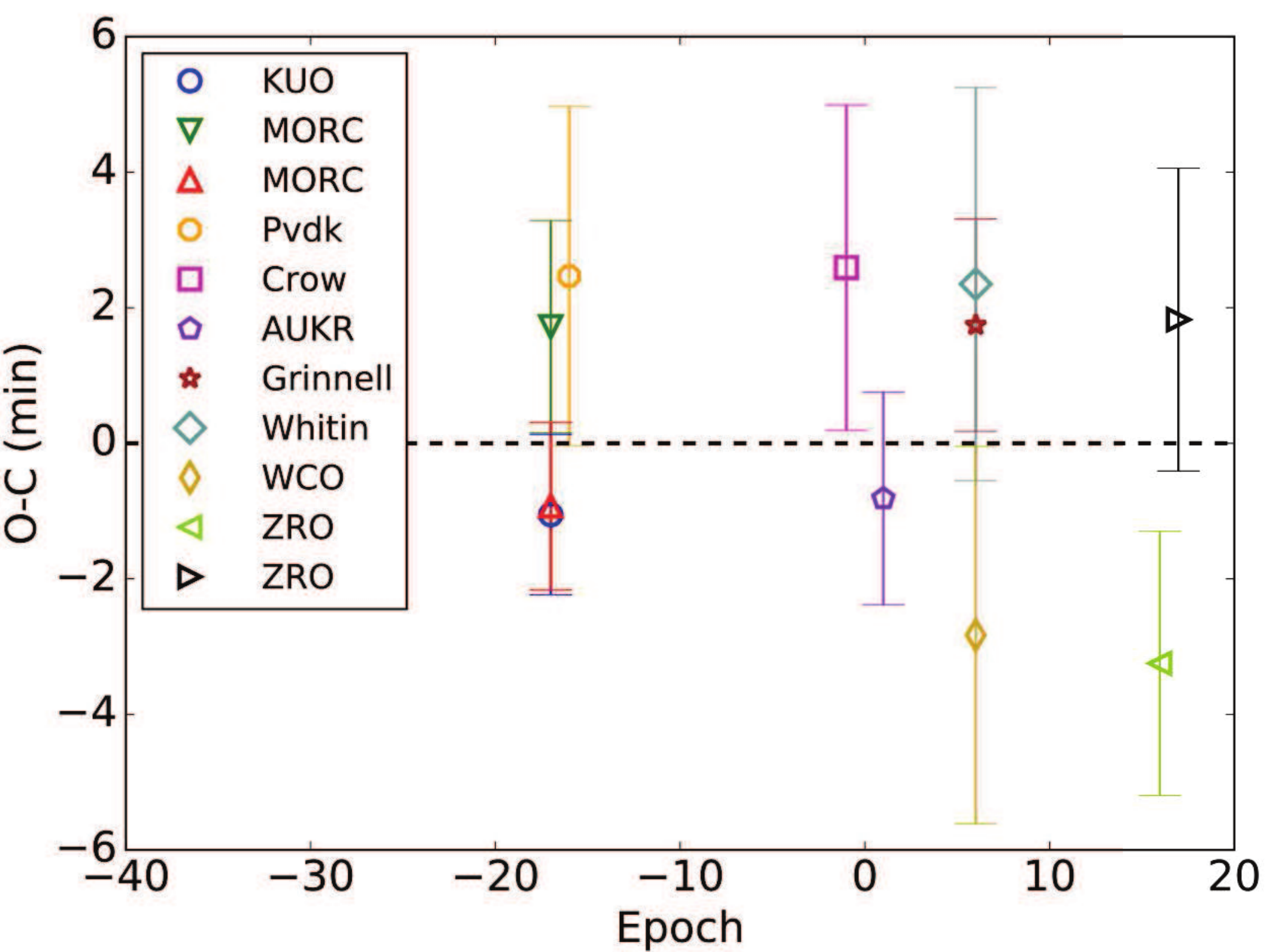}
\caption{\footnotesize Transit time residuals for KELT-18b.  The epoch is given in number of orbital periods relative to the inferior conjunction time from the global fit. The data are listed in Table \ref{tab:TTVs}.  We do not see evidence for TTVs over the relatively short baseline of these observations.}
\label{fig:TTVs}
\end{figure}

To search for possible TTVs that might betray the presence of an additional body in the KELT-18 system, we have computed the observed - computed (O-C) residuals between the mid-transit times determined for the individual light curves and the ones derived from the global fit.  The results are given in Table \ref{tab:TTVs} and plotted in Figure \ref{fig:TTVs}.  The biggest outlier sits only 1.5$\sigma$ away from zero, but that value was derived from a partial light curve that suffered from some residual systematics.  We conclude that we have no evidence for TTVs over the relatively short 3 month baseline of the follow-up photometry.

\begin{table}
\centering
 \caption{Transit times from KELT-18 Photometric Observations\MakeLowercase{b}.}
 \label{tab:TTVs}
 \begin{tabular}{r@{\hspace{12pt}} l r r r c}
    \hline
    \hline
    \multicolumn{1}{c}{Epoch} & \multicolumn{1}{c}{$T_\textrm{C}$} 	& \multicolumn{1}{l}{$\sigma_{T_\textrm{C}}$} 	& \multicolumn{1}{c}{O-C} &  \multicolumn{1}{c}{O-C} 			& Telescope \\
	    & \multicolumn{1}{c}{(\bjdtdb)} 	& \multicolumn{1}{c}{(s)}			& \multicolumn{1}{c}{(s)} &  \multicolumn{1}{c}{($\sigma_{T_\textrm{C}}$)} 	& \\
    \hline
 -17  & 2457493.70451   &  71   &  -63.24   & -0.88  & KUO \\
 -17  & 2457493.7064   &  94   &  103.16   &  1.09 & MORC \\
 -17  & 2457493.70460   &  74   &  -55.64   & -0.75 & MORC \\
 -16  & 2457496.5787   & 150   &  147.98   &  0.99 & Pvdk \\
  -1  & 2457539.6551   & 144   &  155.49   &  1.08 & Crow \\
   1  & 2457545.3962   &  94   &  -48.90   & -0.52 & AUKR \\
   6  & 2457559.7568   &  94   &  104.58   &  1.10 & Grinnell \\
   6  & 2457559.7572   & 174   &  140.87   &  0.81 & Whitin \\
   6  & 2457559.7536   & 167   & -169.74   & -1.01 & WCO \\
  16  & 2457588.4709   & 117   & -194.83   & -1.66 & ZRO \\
  17  & 2457591.3461   & 134   &  109.45   &  0.82 & ZRO \\
    \hline
    \hline
 \end{tabular}
  \begin{flushleft}
  \footnotesize{Epochs are given in orbital periods relative to the value of the inferior conjunction time from the global fit.}
\end{flushleft}
\end{table} 
\subsection{False-Positive Analysis}
\label{sec:False-Positives}

\begin{figure}
\includegraphics[width=1\linewidth]{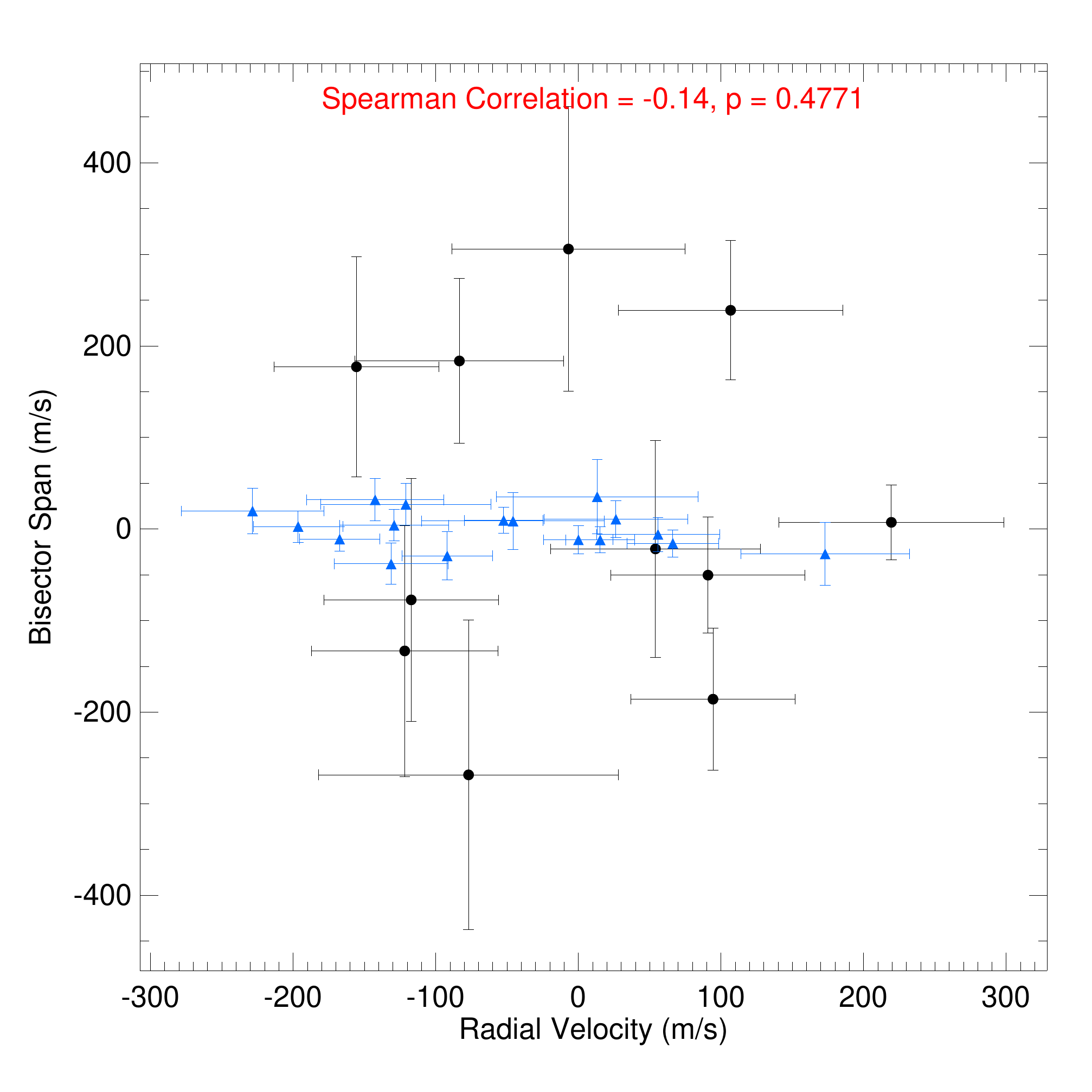}
\vspace{-0.2in}
\caption{\footnotesize Bisector spans for the TRES (blue triangles) and APF (black dots) RV spectra for KELT-18 plotted against the RV values.  We find no correlation between these quantities.}
\label{fig:Bisectors}
\end{figure}

We know that the RV signals are not coming from the faint neighbor because the spectroscopic apertures exclude it.  Several lines of evidence help us to rule out other false-positive scenarios for KELT-18b.  First, our follow-up light curves cover the $g'r'i'z'$ and  $BVRI$ passbands, and are all consistent with the global model as shown in Figure 2. While this evidence is not conclusive on its own, blends often produce detectable light curve depth chromaticity across the optical bands.  Second, examination of a high-resolution spectrum reveals no absorption lines from a second star.  Third, we use the procedures outlined in \citet{Buchave:2010} (TRES) and \citet{Fulton:2015} (APF) to examine the RV bisector spans and check whether the RV variations might instead be caused by spectral line asymmetries due to a nearby eclipsing binary star or stellar activity in KELT-18 itself.  The bisector values are reported in Table \ref{tab:Spectra} and shown in Figure \ref{fig:Bisectors}.   We find a Spearman rank correlation coefficient of $-0.14$ with probability $p=0.477$, giving no indication that the periodic RV signal is due to any astrophysical phenomena other than the orbital motion.  Finally, the AO images rule out any blended source up to 8 mag fainter than KELT-18 at a projected separation of $1\arcsec$ (see Figure \ref{fig:AOContrast}).

None of our radial-velocity observations were obtained during a transit, so we do not have additional information from a Doppler tomographic signal that would add constraints to any blend scenario.

\section{The Neighbor: is it a companion?}
\label{sec:Neighbor}
In a series of recent papers, the Friends of Hot Jupiters collaboration (FOHJ; \cite{Ngo:2016} and references therein) has been examining the frequency, properties, and implications of stellar companions to hot Jupiter hosts.  They find that hot Jupiters are commonly found in multiple-star systems with separations in the range 50-2000 AU with a frequency higher than expected based on the statistics for field stars.  At our derived distance of 311pc (\S\ref{sec:SED}), the projected separation of $3\farcs43 \pm 0\farcs01$ (\S\ref{sec:AO}) between KELT-18 and its on-sky neighbor corresponds to a projected physical separation of $\sim1100\rm AU$, right in this range.  Thus if the two stars are bound, KELT-18 fits the pattern and adds to the collection of  ``friends."  In this section we argue that the neighbor is plausibly a bound companion.  

\subsection{Fluxes}

If the neighbor is a physical companion, we can assume it has the same distance and reddening as KELT-18 and compare its absolute $K_S$ magnitude to the predictions of the \citet{Baraffe:2015} stellar evolutionary models for low-mass stars.  We take the neighbor's apparent magnitude $K_S=12.9 \pm 0.2$ (\S\ref{sec:AO}) and distance $311\pm14$ pc (\S\ref{sec:SED}) to get an absolute magnitude $M_{K_S}=5.44\pm0.22$ (the extinction in $K_S$ is negligible).  For the estimated age of $\sim2$ Gyr (\S\ref{sec:Evol}), the models predict $\rm \teff \sim3900\ \rm K$ which is consistent with our estimate of $\teff$ determined from $K-$ and $z-$band magnitudes in \S\ref{sec:SED}.  This implies that the neighbor is plausibly at the same distance as KELT-18.

\subsection{Sky density}

To investigate this further, we compute the probability that a star of similar or brighter magnitude to the neighbor would be found within $3\farcs43$ of any random point of sky in this region.  We used ds9\footnote{http://ds9.si.edu} to download a 2MASS K-band image surrounding KELT-18 and via its catalog tool determined that in a $1^\circ\times1^\circ$ box there are 
186 (or 268) objects per square degree with K$<$12.9 (or 13.5, which is the 3 sigma faint limit).
This is a relatively small on-sky density because KELT-18 is at high galactic latitude ($b=54^\circ$).  Even for the fainter magnitude, the probability of a chance alignment within any $r=3\farcs43$ circle in this area is only $\sim 0.0008$ for a probability of 0.08\%.  We conclude that with $>3\sigma$ confidence the neighbor is likely a physical companion to KELT-18. 

\subsection{Astrometry and proper motion}

If the neighbor is a bound companion as we suspect, the projected angular separation implies a minimum circular orbital period of $\sim26,000$ years, too long to detect.  Thus, it should effectively travel across the sky with the same proper motion as KELT-18 and the separation should remain constant.  To check for this, we attempted to measure multi-epoch astrometry.  One epoch was provided by our own AO image.  We also measured the separation using the SDSS {\it z}-band image (observed 2001 May), where the neighbor is best-resolved.
To do so, we had to pinpoint the center of KELT-18, which is complicated by its saturation.  We did this using three different metrics: (i) we traced the diffraction spikes across 50\arcsec~ in the two perpendicular directions and found their intersection; (ii) we generated the circular contour just outside the region of saturation at a radius of 2\arcsec; and (iii) we generated a circular contour in the wings of the PSF beyond the distance at which the companion would interfere, at a radius of 8\arcsec.  The three techniques yielded separations of $3\farcs54$, $3\farcs39$, and $3\farcs33$ respectively, with a mean of $3\farcs42$.  The separation and position angle are both in agreement with those determined from the AO measurement.  KELT-18's proper motion of $21\rm mas~yr^{-1}$ (derived from values in Table \ref{tab:LitProps}) would translate to a relative shift of $0\farcs31$ over the 15 yr baseline \textit{if} the neighbor had zero proper motion.  Given the uncertainty in the SDSS measurement we could expect at best a $\lesssim 3\sigma$ detection of a shift, so the fact that we see none is not yet significant.  The much better $0\farcs01$ precision of the AO observations does however suggest that, unless the neighbor has proper motion similar to KELT-18's, a second AO observation with tolerance 0\farcs01 could detect relative motion in just a few years.

\section{Discussion}
\label{sec:Discussion}

\subsection{Comparative Planetology}
KELT-18b is a highly inflated hot Jupiter in a $0.04$ AU circular orbit transiting a 2 Gyr old F4V host star.  Among the known planet hosts, KELT-18 joins a small group that are as hot ($\teff\geq 6600$K), as massive ($M_* \geq 1.5\msun$), and as bright ($V \leq 10.5$)\footnote{NASA Exoplanet Archive (http://exoplanetarchive.ipac.caltech.edu) accessed 2016 Dec 15}.  The other hosts in this group are: HAT-P-49, HAT-P-57, KELT-7, KELT-17, WASP-33, and the extremely massive Kepler-13b ($M_{\rm P}>9\mj$; \citealt{Esteves:2015}).  Of these, KELT-18 hosts the planet with the lowest mass ($1.18\pm0.11\mj$) but also the largest radius ($1.57\pm0.04\rj$), i.e. KELT-18b has the lowest density ($0.375\pm0.04\densitycgs$) among the planets with hot, bright hosts.  Comparing KELT-18b with those planets will help to inform our overall understanding of planet formation around massive stars.

\subsection{Radius inflation}
\begin{figure}
\includegraphics[width=1\linewidth]{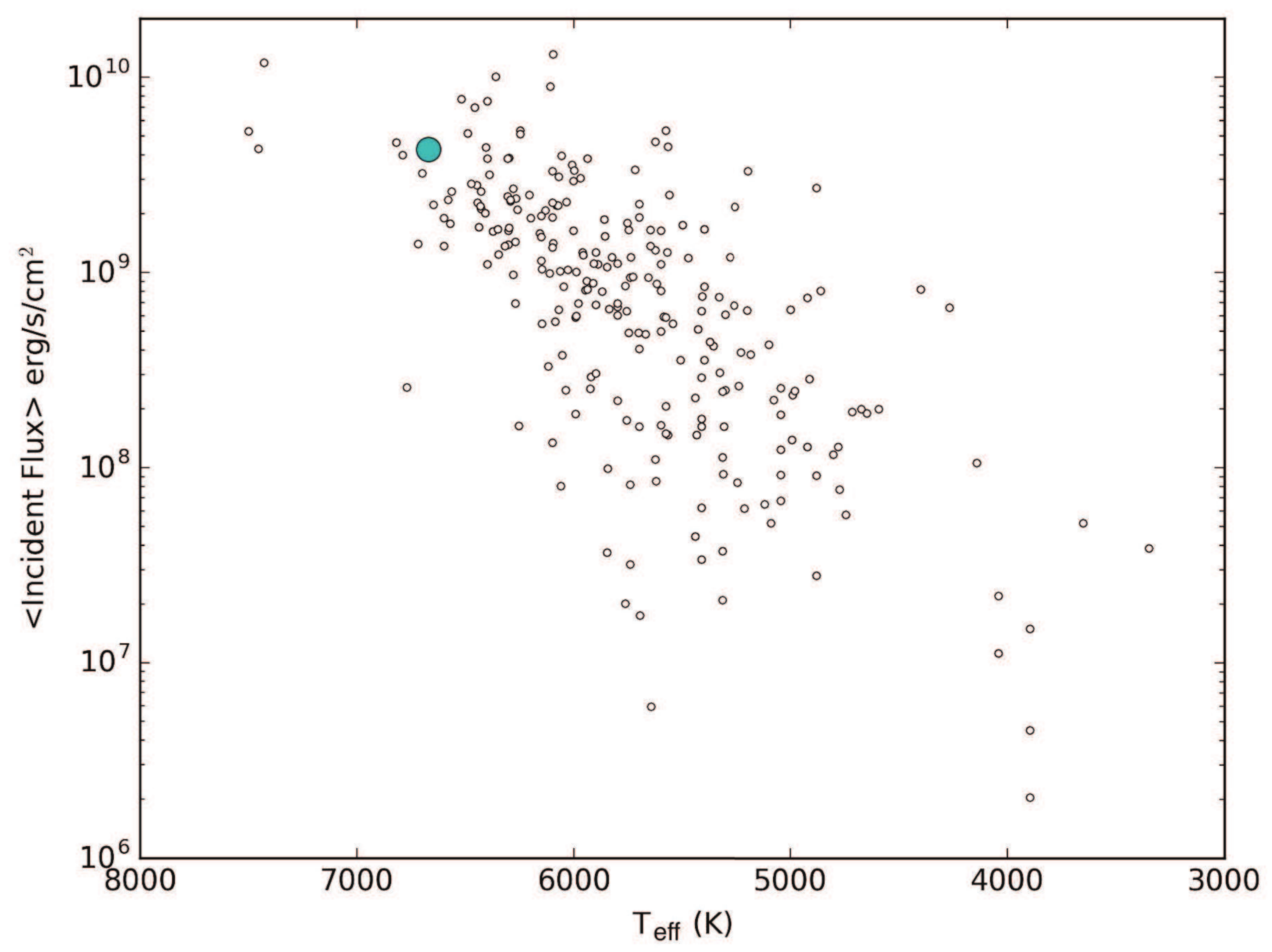}
\caption{\footnotesize Insolation and stellar \teff~ for known exoplanets.  KELT-18b is shown with a large filled circle; its position near the extremes of the distribution make it potentially useful for testing models of the mechanisms of radius inflation.  Based on data from the NASA Exoplanet Archive (http://exoplanetarchive.ipac.caltech.edu) accessed 2017 Jan 12.}
\label{fig:Insolation}
\end{figure}

KELT-18b is large for its mass.  \citet{Chen:2017} have recently compiled a large set of planet radii and masses ($>300$ objects, mostly Jovian worlds,
many of which are inflated planets drawn from ground-based transit surveys) and used them to build a probabilistic model of the relation between them.
Using their {\textit Forecaster}\footnote{https://github.com/chenjj2/forecaster} code we find KELT-18b's radius to be in the upper $\sim8\%$ of those expected for
planets with mass in the same range.  KELT-18's low density is not surprising given that its proximity to its hot host subjects it to a very high level of incident flux, or insolation.   As shown in Table \ref{tab:KELT-18b_global_fit_properties}, the current level is $4.29_{-0.30}^{+0.33} \times\fluxcgs$, which is $\sim20\times$ higher than the threshold for radius inflation \citep{Demory:2011}. KELT-18b's insolation and radius are among the largest for known planets, and it is near an extreme in the parameter space of insolation and host \teff~ as shown in Fig. \ref{fig:Insolation}.  As such, it adds to the collection of objects that can be used to probe the mechanisms and timescales of radius inflation. 

\citet{Weiss2013} have derived an empirical relation for a planet's radius as a function of flux  $F$.  For large planets, defined as $M_P>150 M_{\oplus}\sim0.5\mj$, they find that the insolation is a bigger factor than the mass in determining the radius, and that $R_P/R_{\oplus} = 2.45 (M_P/M_{\oplus})^{-0.039}(F/\rm erg~s^{-1}cm^{-2})^{0.094}$ with an rms scatter of $1.15R_{\oplus}=0.109\rj$.  However, there are only a few planets in the compilation with insolation as high as KELT-18b's, so its addition to the collection adds a useful check.  For KELT-18b, the relation predicts a radius of $R=15.6R_{\oplus}=1.49\pm0.109\rj$, which is consistent with our inferred radius of $1.57\pm0.04\rj$. 

KELT-18b is also consistent with the recent results of \citet{Hartman:2016}, who analyzed hot Jupiter masses and radii together with the evolutionary states of their hosts.  They interpret a relationship between the planetary radius and the stellar fractional age to indicate that hot Jupiters are reinflated as their hosts age through their main sequence lifetimes.  According to their formalism, we find that KELT-18's fractional age (0.6) and KELT-18b's radius put this system right in the middle of their distribution.

\subsection{Potential for atmospheric characterization}

\begin{figure}
\includegraphics[width=1\linewidth]{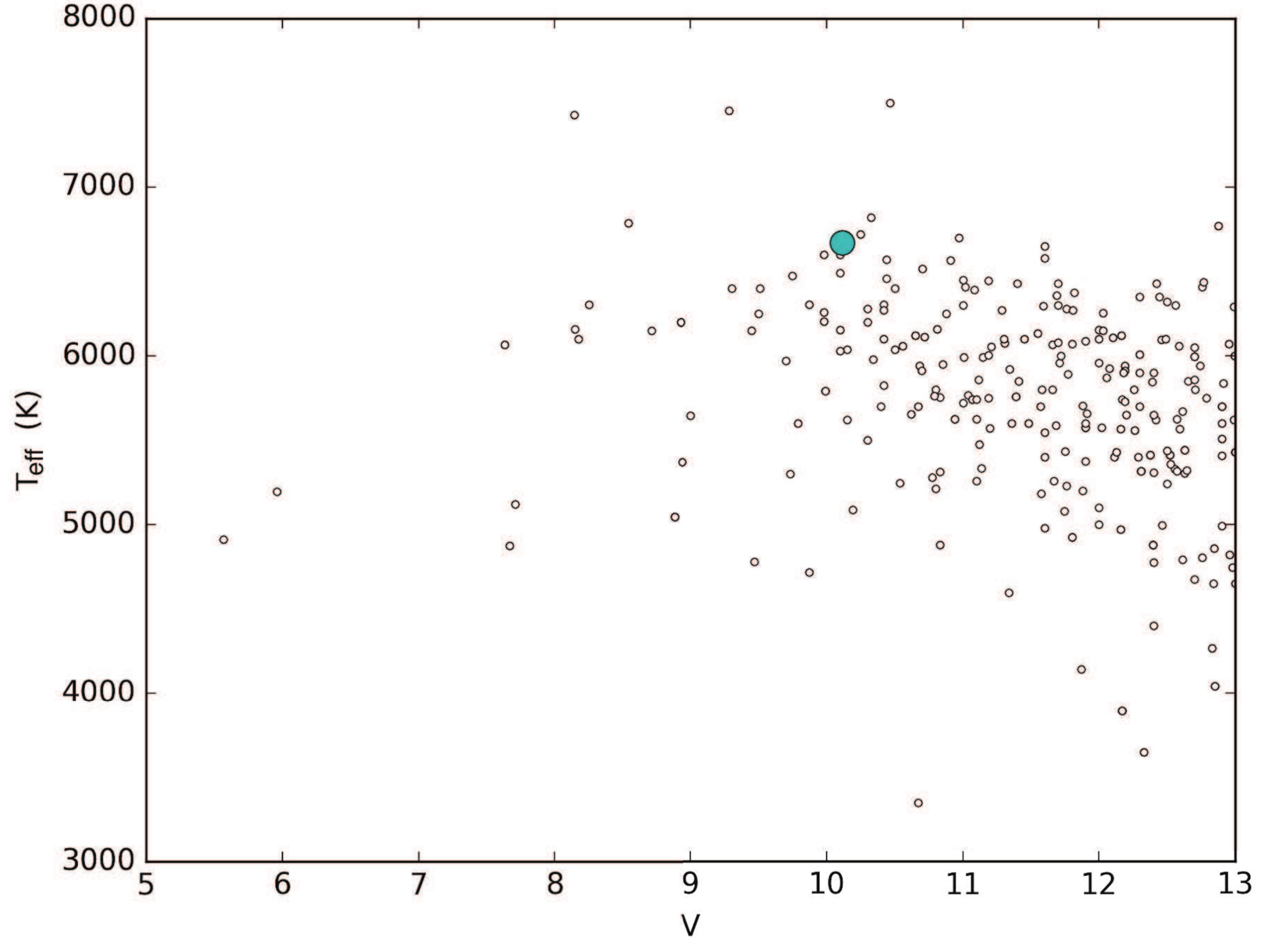}
\caption{\footnotesize Host star \teff~ and $V$-band magnitude for known transiting exoplanets.  KELT-18, shown with a large filled circle, provides an excellent backdrop for atmospheric transmission spectroscopy and infrared photometry during secondary eclipse due to its bright host and low density.  Based on data from the NASA Exoplanet Archive (http://exoplanetarchive.ipac.caltech.edu) accessed 2017 Jan 12.}
\label{fig:Vmags}
\end{figure}

KELT-18b presents an excellent opportunity for observations aimed at atmospheric characterization.  As shown in Fig. \ref{fig:Vmags}, it has a host that is one of the hottest among the brightest hosts of transiting hot Jupiters, much like its southern cousin KELT-14b.  \citet{Rodriguez:2016} describe how KELT-14b's very high equilibrium temperature (1904 K) and bright host star $K$-band magnitude ($K$ = 9.424), make it a prime target for direct detection of thermal emission from the daytime side of the planet through infrared photometric measurements made near secondary eclipse.  KELT-18b provides an even stronger opportunity: the host is even brighter (K = 9.21) and the planet hotter (2100 K).

KELT-18b is also an excellent candidate for atmospheric transmission spectroscopy; it is much like the collection of planets recently observed with the Hubble Space Telescope by \citet{Sing:2016}.  Because the atmospheric scale height $H$ varies inversely with the surface gravity, KELT-18b's low $\log{g_{\rm P}}$~means that features could be relatively strong.  To estimate $H$ we adopt the equilibrium surface temperature $T_{\rm eq}\sim2100$ K and surface gravity $\log{g_{\rm P}}\sim3.07$ \gravitycgs~ from Table \ref{tab:KELT-18b_global_fit_properties} and assume a fiducial mean molecular weight of $\mu=2.3$ to get $H \sim kT_{\rm eq} / (\mu  m_{\rm H}  g_{\rm P}) \sim 600$ km.  For KELT-18b that corresponds to a fractional difference in transit depth with wavelength of up to $\sim2H/R_{\rm P} \sim 1\%$.  KELT-18b could also help to constrain cloud and haze formation scenarios at high temperatures.
An added bonus is that its very hot host means there would be fewer complications from stellar absorption features compared to many of the other bright hosts with later spectral types.  We strongly encourage transmission spectroscopic observations.

\subsection{Spin-orbit misalignment}
\label{sec:SpinOrbit}

KELT-18's effective temperature places it well above the Kraft break \citep{Kraft:1967}, and in the regime where stars are generally rapidly rotating.
The underlying distribution of rotational speeds for main sequence stars as hot as $\teff=$ 6670 K is not well-enough constrained observationally to permit a precise calculation of the inclination based on the observed \vsinistar.  However the recent models of \citet{vanSaders:2013} indicate that for stars with KELT-18's temperature and surface gravity, the rotational velocities are typically in excess of 100 \kms.  The observed slow $\vsinistar=12.3\kms$ thus makes it possible that we are seeing the star close to pole on.  We are led to a similar conclusion from the recent compilation of rotational periods for 24,000 Kepler stars by \citet{Reinhold:2013}, which indicates that for stars in KELT-18's effective temperature range, the distribution of rotational periods is strongly peaked at $P<2$ days. KELT-18's rotational period assuming an edge-on view of the rotation would be $\sim 8$ days, out on the very low-amplitude tail of the distribution.  We conclude that the evidence is suggestive, but not conclusive, that KELT-18 is seen close to pole-on.

\begin{figure}
\includegraphics[width=1\linewidth]{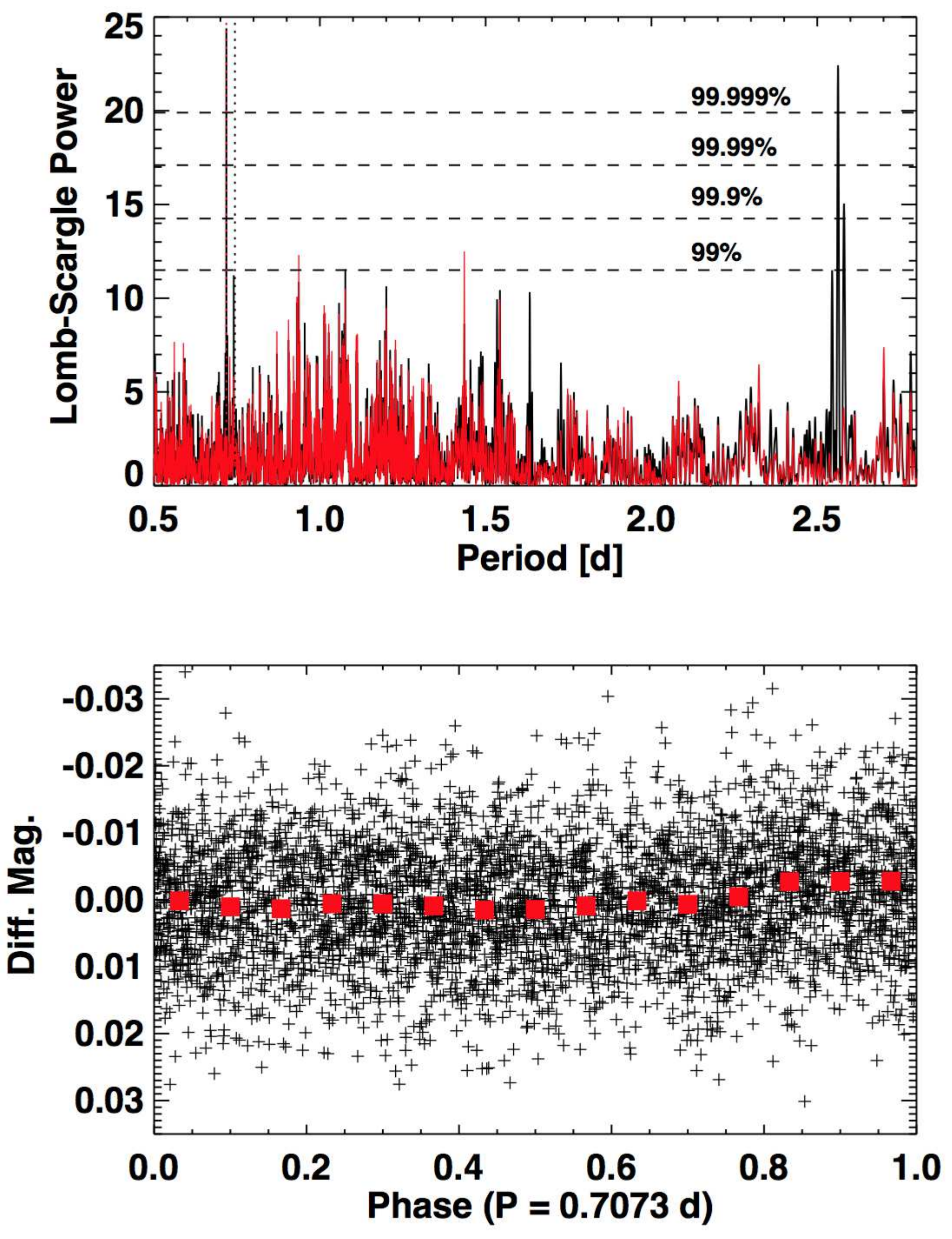}
\caption{\footnotesize KELT-18 Lomb-Scargle analysis.
Top: Lomb-Scargle periodogram of the KELT photometry with in-transit data removed shown in black, and the KELT photometry after filtering out the dominant 0.707-day period shown in red. 
Note that the other strong peaks at $\sim2.5$ days disappear, indicating that they were aliases of the dominant 0.707-day period.  The 0.707 day period (red vertical line) is close to the 1/4 orbital period (black dotted line) but clearly distinct.  
Dashed lines indicate confidence levels at three different values of the power, based on a Monte Carlo determination of the false-alarm
probabilities using a scrambling of the real light curve data. The 0.707-day peak has a false-alarm probability of $<10^{-5}$ (confidence of $>$ 99.999\%).
Bottom: Phased light curve on the 0.707-day period with KELT photometry shown as black crosses.  Red boxes show the same data binned to illustrate the variation more clearly.
}
\label{fig:Rotation}
\end{figure}

In some cases, it is possible to get an independent measurement of the star's rotational speed by carrying out an analysis of time-series photometry to search for periodic signatures such as those that could result from starspot modulation.  KELT-18 is hot enough that these signatures may be weak, but we can check for them using the KELT photometry by generating a Lomb-Scargle \citep{Lomb:1976,Scargle:1982} periodogram.  To do so we start with the KELT photometry and remove the in-transit data.  The resulting periodogram is shown in Fig. \ref{fig:Rotation}.  We detect a signal with a period of 0.707 days and a false-alarm probability of $<10^{-6}$.  There are also slightly smaller peaks near 2.5 days, but these disappear when we filter out the 0.707 day period, indicating that they were aliases of the dominant peak. From a light curve phase-folded on this period (also shown in Fig. \ref{fig:Rotation}) we see a variation with a semi-amplitude of $\sim0.2\%$.  If the 0.707 day period represents the rotational period of the star, then the corresponding equatorial rotation speed would be 134 \kms.  For the observed \vsinistar, this implies an inclination of 5\arcdeg.

Our evidence is suggestive but not yet conclusive that KELT-18 is seen nearly pole-on.  However, if this is the case, then its spin and the orbit of its transiting planet would necessarily be misaligned.
 \citet{Schlaufman:2010} have found that misaligned systems tend to occur for hosts more massive than 1.2\msun.  A misaligned KELT-18 would add to this collection, with a host more massive than any in that study.  It would also fit into the framework proposed by \citet{Winn:2010} and further developed by \citet{Albrecht:2012} based on Rossiter-McLaughlin determinations of projected spin-orbit angles, or obliquities: hot Jupiters are formed with a range of obliquities, which are damped by tides only for the case of hosts with relatively large convective stellar envelopes.  This can include both zero age main sequence $\teff<6250\rm K$ stars and hotter stars once they are old enough to be evolving off the main sequence.  KELT-18's temperature, mass ratio, and orbit size are in the ranges for which high projected obliquities are found.  Future spectroscopic observations during transit should allow an independent check on the alignment for the KELT-18 system, adding a useful high, $\teff\sim6700\rm K$ data point.

One possibility is that KELT-18's high inferred obliquity is related to the presence of its suspected companion (\S\ref{sec:Neighbor}), for example via Kozai-Lidov migration (e.g. \cite{Fabrycky:2007}).  
The FOHJ collaboration \cite{Ngo:2016} and references therein concluded that for hot Jupiters detected by ground-based surveys like KELT, Kozai-Lidov oscillations cannot be the dominant migration mechanism.
Nonetheless we can explore the possibility for KELT-18.  We compute and set equal the Kozai and general relativistic precessional periods (using the \citet{Fabrycky:2007} equations 1 and 23 with eccenticity 0.5 for the stellar orbits following \citet{Ngo:2015}) to find that if KELT-18's neighbor is bound, and its projected separation is the true separation, then the Kozai mechanism could be effective for a formation distance of $\gtrsim5\rm AU$. Thus it is at least plausible as a contributor to the orbital evolution.

\subsection{Tidal Evolution and Insolation History}
\label{sec:Insolation}

\begin{figure}
\includegraphics[width=1\linewidth]{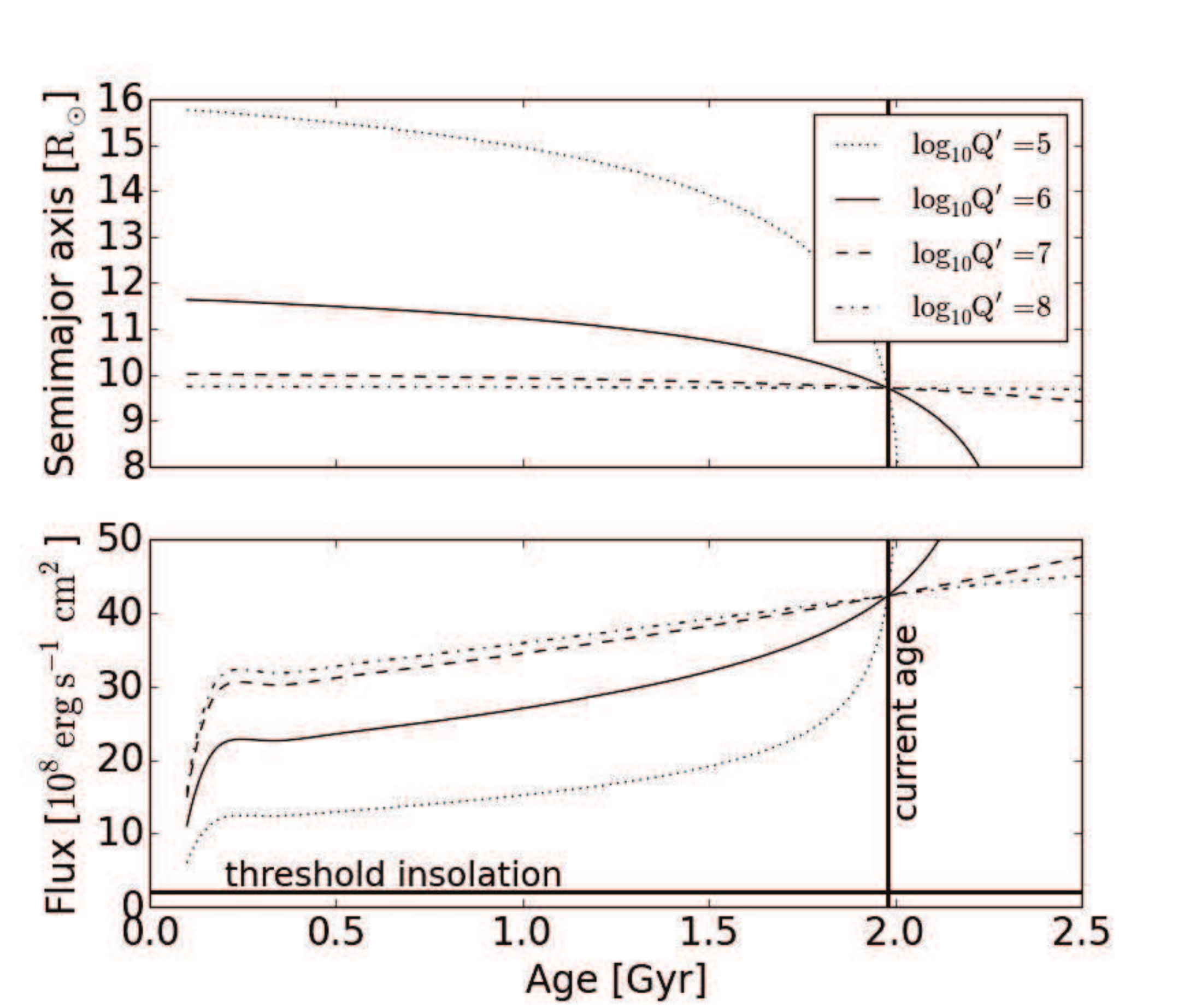}
\caption{\footnotesize Models of the orbital evolution of KELT-18b for a range of stellar tidal quality factors $Q^{\prime}_*$ and  assuming that the star's spin period is longer than the orbital period (which may not be the case; \S\ref{sec:SpinOrbit}).  Top: semimajor axis.  Bottom: insolation.  The horizontal line represents the threshold value of $\rm 2\times10^8 erg\ s^{-1}\ cm^{-2}$ for radius inflation from \citet{Demory:2011}.  The vertical line shows the current age of KELT-18.}
\label{fig:OrbitalEvol}
\end{figure}

If KELT-18 is \textit{not} in fact seen at high inclination but is instead a naturally slow rotator, the measured $\vsinistar$ implies a rotation period of $\sim 8\rm d$.  In this case, we can model the insolation history and future of KELT-18b using the techniques of \citet{Penev:2014} and following the approach described in \citet{Oberst:2016} and \citet{Stevens:2016}.  Briefly, we assume that the host star rotates as a solid body with period longer than the planet's orbit, and that tidal torques (with constant phase lag) exerted by the planet are the only physical influence on the stellar rotation.  We take as boundary conditions the current stellar parameters and orbital semi-major axis from Table \ref{tab:KELT-18b_global_fit_properties} and adopt the appropriate YY stellar evolutionary track to account for the star's changing radius and luminosity with age.  We consider a range of stellar tidal quality factors $Q^{\prime}_*$ where $Q^{\prime -1}_{*}$ is a product of the Love number and the tidal phase lag.  
The results are shown in Figure \ref{fig:OrbitalEvol}.  Assuming that the evolution has been driven by tides alone, we see that KELT-18b's insolation has been well above the radius inflation threshold for the whole main sequence life of its host independent of $Q^{\prime}_*$.  Though other mechanisms (e.g. disk migration, scattering) would have had to bring KELT-18 close to the star initially, we see that for small  $Q^{\prime}_*\sim10^5$, the inward migration due to tides alone could have begun with the planet as much as 60\% farther away than it currently orbits (about 5 stellar radii) and could end as soon as 40 Myr from now as the planet converges on the star.  However, we reiterate that this model is only valid if the star is rotating sub-synchronously, which we believe is unlikely to be the case.

\section{Conclusion}
KELT-18b is a highly inflated hot Jupiter orbiting a hot, F4V star in a 2.87d period.  The host star is very bright ($V=10.1$, $K=9.21$ mag) making this system an excellent candidate for follow-up observations. KELT-18b is one of least dense planets known among those with hot, bright hosts.  It provides a check on the empirical relations for radius inflation in a part of parameter space that is still only sparsely sampled.  KELT-18 has a probable stellar companion at a projected separation of 1100 AU, which may have contributed to the strong misalignment we suspect between KELT-18's spin axis and its planet's orbital axis.  It should be straightforward to verify the companion's status through second-epoch AO imaging and to further explore the spin-orbit alignment through RM or Doppler tomographic measurements.  KELT-18b should be a prime target for atmospheric characterization observations; we strongly encourage follow-up for transmission spectroscopy.

\section{Acknowledgements}
K.K.M. acknowledges the support of the Theodore Dunham, Jr. Fund for Astronomical Research and the NASA Massachusetts Space Grant consortium.  
Work performed by J.E.R. was supported by the Harvard Future Faculty Leaders Postdoctoral fellowship.
D.J.S and B.S.G. were partially supported by NSF CAREER Grant AST-1056524.
KP acknowledges support from NASA grant NNX13AQ62G.
N.N. acknowledges support by a Grant-in-Aid for Scientific Research (A) (JSPS KAKENHI Grant Number 25247026).
OB would like to acknowledge the support by the research fund of Ankara University (BAP) through the project 13B4240006.
EMRK, AB, and YSZ were supported by the Research Corporation for Science Advancement through the Cottrell College program.  AB and YSZ acknowledge funding from the Grinnell College Mentored Advanced Project (MAP) program.
We acknowledge Noriyuki Matsunaga for providing time with the Subaru IRCS. 

This work has made use of NASA’s Astrophysics Data System, the Extrasolar Planet Encyclopedia at exoplanet.eu, the SIMBAD database operated at CDS, Strasbourg, France, and the VizieR catalogue access tool, CDS, Strasbourg, France.  We also used data products from the Widefield Infrared Survey Explorer, which is a joint project of the University of California, Los Angeles; the Jet Propulsion Laboratory/California Institute of Technology, which is funded by the National Aeronautics and Space Administration; the Two Micron All Sky Survey, which is a joint project of the University of Massachusetts and the Infrared Processing and Analysis Center/California Institute of Technology, funded by the National Aeronautics and Space Administration and the National Science Foundation; the American Association of Variable Star Observers (AAVSO) Photometric All-Sky Survey (APASS), whose funding is provided by the Robert Martin Ayers Sciences Fund and the AAVSO Endowment (\url{https://www.aavso.org/aavso-photometric-all-sky-survey-data-release-1}); and the European Space Agency (ESA) mission {\it Gaia} (\url{http://www.cosmos.esa.int/gaia}), processed by the {\it Gaia} Data Processing and Analysis Consortium (DPAC, \url{http://www.cosmos.esa.int/web/gaia/dpac/consortium}). Funding for the DPAC has been provided by national institutions, in particular the institutions participating in the {\it Gaia} Multilateral Agreement.

Facilities:
\facility{KELT (North)},
\facility{APF (Levy)},
\facility{Keck:I (HIRES)},
\facility{FLWO:1.5m (TRES)},
\facility{Subaru (IRCS)},
\facility{WCWO:0.6m}
\facility{MO:0.6m}

\end{document}